\documentclass[aps,pre,superscriptaddress,amsmath,amssymb,twocolumn]{revtex4}
\usepackage{graphicx}
\usepackage{hyperref}
\usepackage{pifont}
\usepackage{color}
\definecolor{darkblue}{rgb}{0.0,0.0,0.5}
\hypersetup{colorlinks,breaklinks,
            linkcolor=darkblue,urlcolor=darkblue,
           anchorcolor=darkblue,citecolor=darkblue}

\begin{document}

\title{The Effect of Substrate on Thermodynamic and Kinetic Anisotropies in Atomic Thin Films}

\author{Amir Haji-Akbari}
\affiliation{Department of Chemical and Biological Engineering, Princeton University, Princeton NJ 08544}

\author{Pablo G. Debenedetti}
\email{pdebene@exchange.princeton.edu}
\affiliation{Department of Chemical and Biological Engineering, Princeton University, Princeton NJ 08544}

\date{\today}

\begin{abstract}
Glasses have a wide range of technological applications. The recent discovery of ultrastable glasses that are obtained by depositing the vapor of a glass-forming liquid onto the surface of a cold substrate has sparked renewed interest in the effects of confinements on physicochemical properties of liquids and glasses. Here we use molecular dynamics simulations to study the effect of substrate on thin films of a model glass-forming liquid, the Kob-Andersen binary Lennard-Jones system, and compute profiles of several thermodynamic and kinetic properties across the film. We observe that the substrate can induce large oscillations in profiles of thermodynamic properties such as density, composition and stress, and we establish a correlation between the oscillations in total density and the oscillations in normal stress. We also demonstrate that the kinetic properties of an atomic film can be readily tuned by changing the strength of interactions between the substrate and the liquid. Most notably, we show that a weakly attractive substrate can induce the emergence of a highly mobile region in its vicinity. In this highly mobile region, structural relaxation is several times faster than in the bulk, and the exploration of the potential energy landscape is also more efficient. In the subsurface region near a strongly attractive substrate, however, the dynamics is decelerated and the sampling of the potential energy landscape becomes less efficient than the bulk. We explain these two distinct behaviors by establishing a correlation between the oscillations in kinetic properties and the oscillations in lateral stress. Our findings offer interesting opportunities for designing better substrates for the vapor deposition process or developing alternative procedures for situations where vapor deposition is not feasible.
\end{abstract}

\maketitle

\section{Introduction\label{section:intro}}

A major technological trend of the last few decades has been the shrinkage of accessible time and length scales. This has made precision manufacturing of nano-sized particles and devices easier than ever before. An important consequence of this trend is the famous "Moore's Law" which states that the computing power of a transistor chip doubles every 18 months~\cite{IEEESpectrum1997}. The physicochemical properties of matter at reduced length scales can deviate significantly from the bulk. These differences can be partly attributed to quantum effects that are only present at such small length scales~\cite{MoiseeviNature1996,DanielChemRev2004,MarinicaNanoLetters2012}, but they can also be a result of confinement. It is thus practically important to understand the effect of confinement on the properties of atomic and molecular systems. Confined states of matter are not only present in the state-of-the-art technologies, but are also ubiquitous in nature. Confinement plays an important role in determining the behavior of systems as diverse as biological cells~\cite{MintonJBC2001,SkolnickPNAS2010}, aquifers~\cite{WangJHydrol2012}, natural gas and oil reservoirs~\cite{ChannelJGR1999} and atmospheric droplets and aerosols that constitute clouds~\cite{WestbrookQJRMS2013}. Such ubiquity endows the study of confinement in materials with unusually broad interest.

\begin{figure}
	\begin{center}
		\includegraphics[width=.33\textwidth]{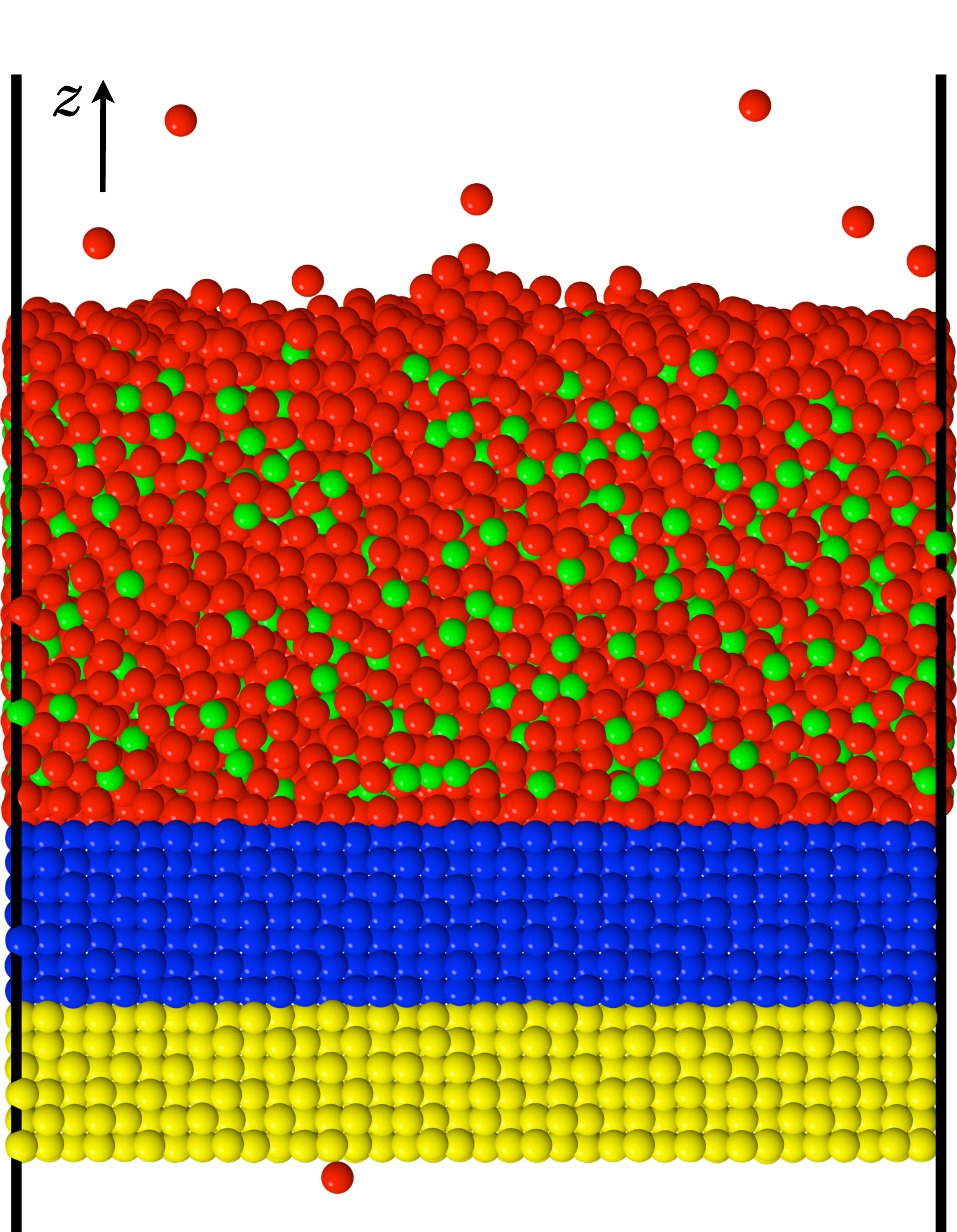}
		\caption{\label{fig:schematic}A schematic representation of the simulation box. The liquid film is made up of red (A) and green (B) atoms, while the blue (C) and yellow (D) atoms comprise the attractive and repulsive substrates respectively.}
	\end{center}
\end{figure}

What is universal about confinement is that it breaks the translational isotropy of a bulk material. As a consequence, all physicochemical properties become functions of position in confined states of matter. It has indeed been shown in various experimental~\cite{IsraelachviliJMatRes1990, SpencerScience2001, BellJACS2003, ZhuEdigerPRL2011, HoffmanPRL2010, ForestScience2014} and computational~\cite{HadziioannouJCP1990, DavisJCP1992, RosskyJCP1994, BerendsenJPhysChem1994, BrinkeLangmuir1995, LandmanPRL1997, FuchsJCP2002, TeboulJPhysCondMat2002, OstrowskyPhysicaA2002, BerneJPhysChemB2004, KumarJCP2005, ShiJCP2011, deBeerEPL2012, StrioloJPCC2012} studies of confined matter that properties such as density, composition, diffusivity and viscosity can be strong functions of position. This lack of translational isotropy can lead to major changes in the global properties of the corresponding material. For instance, confinement can shift  coexistence manifolds in the thermodynamic phase diagram of the bulk system, e.g.~by lowering the melting temperature~\cite{ChristensonJPCM2001,TaschinEPL2010,RichertARPC2010, PriestleyMacromolecules2011,OguniJPCB2011}. It can also induce new phases that are not possible in the bulk~\cite{MolineroJCP2010}, or can alter the effective dynamic and mechanical properties of matter such as relaxation times~\cite{LevingerARAC2010,RichertARPC2010,HoffmanPRL2010}, elastic constants~\cite{HirvonenJAppPhys1997, StaffordSoftMatter2009,HoffmanPRL2010}, viscosities~\cite{BrinkeLangmuir1995,StaffordSoftMatter2009,RafiqJPCB2010} and diffusivities~\cite{BarratEPL1995,deBeerEPL2012,KrishnamoortiACSNano2013}. 

The translational anisotropy of confined systems can be utilized to produce materials with superior properties. Vapor-deposited ultrastable glasses discovered by Swallen~\emph{et al}.~\cite{EdigerScience2007} are notable examples. Unlike ordinary glasses that are prepared via gradual cooling of the \emph{bulk} supercooled liquid~\cite{TuckerJCP1974,DebenedettiNature2001}, ultrastable glasses are obtained by depositing the vapor of the glass forming substance onto the surface of a substrate with a temperature that is slightly below $T_g$. The arising glasses tend to possess extraordinary thermodynamic and kinetic stability (e.g.~higher onset temperature and density). Calorimetric experiments reveal that these glasses have lower enthalpies in comparison to ordinary glasses~\cite{EdigerScience2007}. They thus correspond to configurations that have descended deeper along their potential energy landscape than the corresponding 'bulk` glasses. Although certain structural features of such low-energy amorphous configurations, such as the abundance of regular polyhedral Voronoi cells, have been described in molecular simulations of model atomic stable glasses~\cite{SinghNatMater2013}, considerable gaps in understanding remain regarding the mechanism(s) through which vapor deposition gives rise to such stable configurations. It was argued in the original publication of Swallen~\emph{et al} that the formation of stable glasses is facilitated by the existence of a highly mobile region in the growing free interface, which is a direct consequence of translational symmetry breaking in the supercooled liquid.  Self-diffusivity measurements in organic glasses around the glass transition temperature ($T_g$) have indeed revealed the existence of a highly mobile region in the vicinity of the free surface, with diffusivities as large as $10^6$ times the bulk self-diffusivity~\cite{ZhuEdigerPRL2011}. A similar behavior has been observed in viscosity measurements for glassy 3-methylpentane films~\cite{BellJACS2003}. Also, glassy polymer films have been observed to have a highly mobile region in the vapor-liquid interface at temperatures below $T_g$~\cite{ForestScience2014}. The existence of such a high mobility region has been observed in molecular simulations of free-standing atomic thin films as well~\cite{ShiJCP2011}. It has also been observed that the melting of these stable glasses proceeds via an interface-initiated growth front that develops at interfaces and grows deep into the bulk amorphous solid~\cite{SwallenPRL2009}. The original stable glasses of Swallen~\emph{et al} were prepared from indomethacin and 1,3-bis-(1-naphthyl)-5-(2-naphthyl)benzene. The same procedure has been used to prepare stable glasses of other organic molecules such as ethylbenzene~\cite{IshiiCPL2008, GutierrezPCCP2010}, tuloen~\cite{GutierrezTA2009, GutierrezJPCL2010, GutierrezPCCP2010,SepulvedaPRL2011},  isopropyl benzene~\cite{IshiiJPCB2012}, decalin~\cite{WhitakerJPCB2013} and 1-propanol~\cite{SoudaJPCB2010}. Also, similar procedures have been used to prepare ultrastable polymeric~\cite{PriestleyNatMat2012} and metallic glasses~\cite{SamwerAdvMat2013}.  It has thus been suggested that the formation of stable glasses via vapor deposition is a universal process that can be used for preparing stable glasses from any glass-forming substance~\cite{ZhuCPL2010}.

Glasses have been of considerable technological interest even before the discovery of ultrastable glasses, thanks to their high level of structural uniformity, with widespread applications in areas such as microfluidics~\cite{SukasLabChip2013}, energy storage~\cite{MasoudJAC2013, ZardettoNanotech2013}, medicine~\cite{ElliottJNCS2003,RegiEJIC2003,FurtosPST2013}, photonics and laser technology~\cite{GapontsevOLT1982, GuoJLum2013, GhoshCPL2013}, lithography~\cite{CarapellaJNCS2013}, spectroscopy~\cite{LeeMRB2013}, tissue engineering~\cite{OliveiraBM2013} and chromatography~\cite{WeetaliME1974}. In most technological applications of glasses, the behavior of the glassy component is strongly affected by the presence of one or more interfaces. Because of this, and because of the potential role of confinement in promoting superior stability in vapor-deposited glasses, it is crucial to study translational anisotropy in atomic and molecular thin films. Computer simulations are invaluable tools in this endeavor since structural and dynamical heterogeneities that develop due to lack of translational isotropy can be probed with much higher resolution in simulations than in experiments. 

In this work, we use molecular dynamics simulation in order to perform a systematic investigation of thermodynamic and kinetic anisotropies in thin films of a model atomic glass-forming liquid, the Kob-Andersen binary Lennard-Jones system~\cite{KobAndersenPRE1995}. This paper is organized as follows. We describe the particulars of the system studied in this work in Section~\ref{methods:system}. Technical specifications of molecular dynamics simulations as well as the procedure used for system preparation are presented in Section~\ref{methods:preparation}. In Section~\ref{methods:profiles}, we describe the numerical procedures used for computing position-dependent thermodynamic and kinetic properties. We present and discuss the computed profiles of those thermodynamic and kinetic properties in Sections~\ref{results:static} and~\ref{results:dynamic} respectively. And finally, Section~\ref{discussion} is reserved for concluding remarks. 

\section{Methods}

\subsection{System Description and Interatomic Potential \label{methods:system}}
Fig.~\ref{fig:schematic} shows a schematic representation of the simulation box that contains a liquid film (depicted in red and green), an attractive substrate (depicted in blue) and a repulsive substrate (depicted in yellow). The box is a periodic cuboidal cell stretched along the $z$ direction in order to avoid the film and the substrates from being affected by their periodic images. The system is made up of four types of atoms (or particles) that have identical masses and interact via the Lennard-Jones (LJ) potential~\cite{LJProcRSoc1924}:
\begin{eqnarray} 
U_{\alpha\beta} (r) &=& 4\epsilon_{\alpha\beta}\left[\left(\frac{\sigma_{\alpha\beta}}r\right)^{12}-\left(\frac{\sigma_{\alpha\beta}}r\right)^{6}\right]
\end{eqnarray}
All LJ potentials are truncated at a cutoff value of $r_{c,\alpha\beta}$ and shifted to zero at $r_{c,\alpha\beta}$. All quantities in this work are expressed in LJ reduced units based on $\epsilon_{AA}$, $\sigma_{AA}$ and $m_A$. The interaction parameters for all pairs and the corresponding cutoff values are given in Table.~\ref{table:LJparam}. The dimensions of the box are given by $L_x=L_y=18\sqrt[3]{4}\sigma_{AA}\approx28.57\sigma_{AA}$ and $L_z=100\sqrt[3]{4}\sigma_{AA}\approx158.74\sigma_{AA}$.

The liquid film in Fig.~\ref{fig:schematic} is made up of a binary mixture of A and B atoms that are depicted in red and green respectively, and are mixed at a 4:1 ratio. The A--A, A--B and B--B interaction parameters correspond to the Kob-Andersen binary LJ mixture~\cite{KobAndersenPRE1995}. 
The binary mixture arising from this particular choice of parameters is a liquid that has never been observed to crystallize or  phase separate under normal circumstances~\cite{UlfJCP2009}, and is therefore a popular model for studying atomic glass-forming liquids.

\begin{table}
\begin{center}
	\caption{\label{table:LJparam}$\epsilon$, $\sigma$ and cutoff values for the Lennard Jones potential.}
	\begin{tabular}{c|c|ccc}
		\hline\hline
		~~~~~$\alpha$~~~~~& ~~~~~$\beta$~~~~~ & ~~~~~$\epsilon_{\alpha\beta}$~~~~~ & ~~~~~$\sigma_{\alpha\beta}$~~~~~ & ~~~~~$r_{c,\alpha\beta}$~~~~ \\
		\hline
		A & A & $1.00$ & $1.00$ & $3.50$ \\
		B & B & $0.50$ & $0.88$ & $3.08$ \\
		C & C & $\epsilon_S$  & $1.00$ & $3.50$ \\
		D & D & $\epsilon_S$  & $1.00$ & $3.50$  \\
		A & B & $1.50$ & $0.80$ & $2.80$ \\
		A & C & $\epsilon_S$  & $1.00$ & $3.50$ \\
		A & D & $\epsilon_S$  & $1.00$ & $1.10$ \\
		B & C & $\epsilon_S$  & $1.00$ & $3.50$ \\
		B & D & $\epsilon_S$  & $1.00$ & $1.10$ \\
		C & D & $\epsilon_S$  & $1.00$ & $3.50$ \\
		\hline
	\end{tabular}
\end{center}
\end{table}

The liquid film is in the vicinity of an attractive substrate that is shown in blue in Fig.~\ref{fig:schematic} and is made up of C atoms that attract A and B atoms with $\epsilon_{AC}=\epsilon_{BC}=\epsilon_S$. Unlike A and B atoms that are mobile during the course of the molecular dynamics simulation, C atoms are located on the sites of a face-centered cubic (fcc) lattice with a reduced number density of $1.00$ and are tethered to their lattice positions using stiff harmonic springs with  spring constant $k_s\sigma_{AA}^2/\epsilon_{AA}=10^3$. In all our simulations, the liquid film is placed close to the [001] face of the fcc crystal. On the opposite side of the attractive substrate there is a repulsive substrate made up of D atoms (shown in yellow in Fig.~\ref{fig:schematic}). The A--D and B--D LJ interactions are truncated and shifted at $r_c=1.1\sigma_{AD}=1.1\sigma_{BD}$ and are therefore purely repulsive. Note that $\epsilon_{AD}=\epsilon_{BD}=\epsilon_S$. D atoms are also placed onto the same fcc lattice occupied by C atoms, and are tethered to their lattice positions with harmonic springs of equal stiffness. If a single attractive substrate is included, a second liquid film might develop on the opposite side of the attractive substrate due to evaporation and re-deposition of A and B atoms as a result of the periodic boundary conditions. By including a second repulsive substrate, we make sure that the thickest possible film is formed and maintained during the course of the simulation.  In order to understand the effect of substrate/liquid interactions, we carried out our simulations for three different values of $\epsilon_S$, namely $0.3, 0.5$ and $1.0$.

\subsection{Molecular Dynamics Simulations and System Preparation
\label{methods:preparation}}
The molecular dynamics (MD) simulations are performed in the $NVT$ ensemble using LAMMPS~\cite{PlimptonJCompPhys1995}. Newton's equations of motion are integrated using the velocity Verlet algorithm~\cite{SwopeJCP1982} with a time step of $\Delta{t}=0.0025$, and the temperature is controlled using the Nos\'{e}-Hoover thermostat~\cite{NoseMolPhys1984,HooverPhysRevA1985} with a time constant of $\tau=2.0$. For each value of $\epsilon_S$, four simulations are performed at the reduced temperatures $T^*=0.6, 0.7, 0.8$ and $0.9$.

The initial configuration of the system is prepared as follows. The attractive and repulsive substrates are constructed by placing C and D atoms at their corresponding lattice positions. A similar fcc crystal comprised of A atoms is constructed in the vicinity of the [001] face of the attractive substrate, and the identity of each A atom is randomly changed to B with a probability of $p=0.2$. The number of atoms used in each of these lattices is given in Table.~\ref{table:CD_count}.  The resulting film is equilibrated in a molecular dynamics simulation in which two separate thermostats are used. All substrate atoms are coupled to a thermostat that is set to the final target temperature of the system. The mobile A and B atoms, however, are coupled to a second thermostat that is initially set to $T^*=0.9$, a high temperature for the Kob-Andersen LJ mixture. During the initial stage of the MD simulation that is performed for $10^5$ MD steps, the binary LJ crystal melts to become  a 'hot` liquid film. The target temperature of the second thermostat is then gradually decreased to the final target temperature at a rate of  $10^{-6}$/MD step. Eventually, the system is equilibrated for $1.5\times10^6$ MD steps with both thermostats set to the final target temperature of the system.  As will become apparent in Section~\ref{results:dynamic}, the initial melting and the final equilibration times are orders of magnitude greater than the largest structural relaxation times in all systems simulated in this work.
The final configurations obtained from the procedure described above are used for production runs that are performed in the $NVT$ ensemble at the final target temperature. We use an in-house C++ computer program that links against the LAMMPS static library to calculate the profiles of thermodynamic and kinetic properties on-the-fly. We sample each trajectory every $m$ MD steps where $m$ depends on temperature and is larger for lower temperatures. The numerical procedures used for calculating these profiles are presented in the following section.

\subsection{\label{methods:profiles}Translational Anisotropy and Profiles of Structural and Dynamical Properties}
\subsubsection{Thermodynamic Properties\label{section:methods:structural}}
In order to study the position dependence of thermodynamic properties, the liquid film is partitioned into slices that are each $\sigma_{AA}/20$ thick, and simple time averages of the corresponding properties are calculated in each slice. For instance, the density profile of atoms of type $\alpha=\text{A},\text{B}$ is given by:
\begin{eqnarray}
\rho_{\alpha}(z) &=& \frac{1}{S^{\parallel}}\left\langle\sum_{i=1}^{N_\alpha}\delta(z_{\alpha,i}-z)\right\rangle
\end{eqnarray}
where $S^{\parallel}=648\sqrt[3]{2}\sigma_{AA}^2$ is the cross-sectional surface area of each slice. To determine the potential energy profile, an energy is assigned to each atom by equally splitting the energy contribution of each pair between the two participating atoms. The potential energy profile is thus given by:
\begin{eqnarray}
U_{\alpha}(z) & = &\frac
{\left\langle\sum_{i=1}^{N_\alpha}U_{\alpha,i}\delta(z_{\alpha,i}-z)\right\rangle}
{\left\langle\sum_{i=1}^{N_\alpha}\delta(z_{\alpha,i}-z)\right\rangle}\label{eq:pot}
\end{eqnarray}
$U_{\alpha}(z)$ is a measure of the \emph{average} potential energy of $\alpha$-type atoms residing in a slice that contain $z$. In order to calculate stress profiles, we use the virial equation~\cite{TestaJCP2006}:
\begin{eqnarray}
\mathcal{S}^{\lambda\nu}(z)&=&-\frac{1}{S^{\parallel}}\Bigg\langle\sum_{i=1}^N\delta(z-z_i)\Bigg(\frac{p_i^\lambda p_i^\nu}{m_i}\notag\\&&+\frac12\sum_{j(\neq i)=1}^Nr_{ij}^\lambda \mathcal{F}_{ij}^\nu(\textbf{r}_{ij})\Bigg)\Bigg\rangle\label{eq:stress}
\end{eqnarray}
where $\lambda,\nu=x,y,z$, $N$ is the total number of atoms in the system, $\textbf{r}_i$ and $\textbf{p}_i$ are the position and the momentum of the $i$th atom, $\textbf{r}_{ij}=\textbf{r}_i-\textbf{r}_j$ and $\mathcal{F}_{ij}(\cdot)$ is the force exerted by the $j$th atom on the $i$th atom.  Note that $\mathcal{S}^{xx}(z)=\mathcal{S}^{yy}(z)\neq \mathcal{S}^{zz}(z)$ due to lack of translational isotropy in the system. In order to make this distinction, we will refer to the $xx$ and $yy$ components of the stress tensor as the \emph{lateral stress} and to its $zz$ component as the \emph{normal stress}. The notion of normal stress here should not be confused with the well-known notion of normal stress in hydrodynamics that is typically used against shear stress. Another structural property that was calculated in this study is the \emph{lateral radial distribution function}, which is the radial average of the pair distribution function:
\begin{eqnarray}
&&g_{\alpha\beta}(\textbf{r}^{\parallel},z)=\notag\\&& \frac{\left\langle
\sum\limits_{i=1}^{N_\alpha}\sum\limits_{j=1}^{N_\beta}
\delta\left(\textbf{r}^{\parallel}_{\alpha,i}-\textbf{r}^{\parallel}_{\beta,j}-\textbf{r}^{\parallel}\right)
\mathfrak{D}(z_{\alpha,i},z_{\beta,j},z)
\right\rangle}{\rho_{\alpha}(z)\rho_{\beta}(z)}\label{eq:rdf_slice}
\end{eqnarray}
where $\textbf{r}^{\parallel}\equiv(x,y)\in\mathbb{R}^2$ is the lateral projection of $\textbf{r}\in\mathbb{R}^3$ onto the $xy$ plane and $\mathfrak{D}(z_1,z_2,z)=\delta(z_1-z)\delta(z_2-z)$. 

We use the MD-based FIRE algorithm~\cite{BitzekPRL2006} to perform energy minimization on all configurations sampled by our in-house program. This algorithm enables us to perform energy minimization despite having an interatomic potential that has a discontinuity in its first derivative. We then obtain the profile of potential energies in the inherent structure (IS) using the following equation: 
\begin{eqnarray}
U_{IS,\alpha}(z) & = &\frac
{\left\langle\sum_{i=1}^{N_\alpha}U_{IS,\alpha,i}\delta(z_{\alpha,i}^0-z)\right\rangle}
{\left\langle\sum_{i=1}^{N_\alpha}\delta(z_{\alpha,i}^0-z)\right\rangle}\label{eq:pot}
\end{eqnarray}
with $z_{\alpha,i}^0$ the $z$ coordinate of particle $i$ (of type $\alpha$) \emph{prior} to energy minimization. 

\subsubsection{Kinetic Properties\label{methods:dynamical}}
There are rigorous expressions that relate kinetic properties, such as transport coefficients, of a closed system to mechanical observables of its microstates.  The most well-known approach-- known as the 'Green-Kubo` approach-- relates transport coefficients to integrals of time autocorrelations of microscopic currents~\cite{GreenJCP1954, KuboJPSJ1957}. A second approach developed by Helfand~\cite{HelfandPhysRev1960} expresses transport coefficients in Einstein-like equations in terms of moments that are time integrals of microscopic currents~\cite{GaspardPRE2003}. But no rigorous equivalent expression exists for open systems since both these approaches rely on autocorrelations that are not trivial to define when particles can move in and out of a control volume. This is particularly problematic in studying dynamical heterogeneities in confined systems-- including the liquid films studied in this work-- since the compartments obtained from partitioning the confined systems (such as the cuboidal slices defined in Section~\ref{section:methods:structural}) are open systems that exchange particles with their neighboring compartments.

\begin{table}
\begin{center}
	\caption{\label{table:CD_count}
	The dimensions of the fcc lattice and the total number of different types of atoms in the initial configuration used for system preparation.
	}
	\begin{tabular}{ccccc}
		\hline\hline
		~~~~~Type~~~~~& ~~~~~$n_x$~~~~~ & ~~~~~$n_y$~~~~~ & ~~~~~$n_z$~~~~~ &~~~~~$N$~~~~~\\
		\hline
		A+B & 18 & 18 & 10 & $12\,960$\\
		C & 18 & 18 & 3.5 & $4\,536$\\
		D & 18 & 18 & 3 & $3\,888$\\
		\hline
	\end{tabular}
\end{center}
\end{table}

This lack of a rigorous formalism for computing transport coefficients in open systems is a limitation that affects all computational studies of confined systems. Although several authors have attempted to develop novel problem-specific rigorous algorithms for computing position-dependent transport coefficients~\cite{Hummer2005,MittalPRL2006}, most have resorted to phenomenological approaches that are based on the existing expressions for closed systems. We describe the existing phenomenological approaches in the context of computing lateral diffusivities in systems that are anisotropic in one dimension only, like the liquid films studied in this work. The problem of anisotropic diffusion is described by the Smoluchowski equation~\cite{SanoJCP1981}:
\begin{eqnarray}
\left\{\frac{\partial}{\partial t}-\nabla\left[e^{-\beta f(\textbf{r})}\textbf{D}(\textbf{r})\cdot\nabla e^{\beta f(\textbf{r})}\right]  \right\}p(\textbf{r},t|\textbf{r}_0,t_0) &=& 0\notag\\ &&\label{eq:smoluchowski}
\end{eqnarray}
where $p(\textbf{r},t|\textbf{r}_0,t_0)$ is the conditional probability of observing a particle at position $\textbf{r}$ at time $t$ if it was located at $\textbf{r}_0$ at time $t_0$, $\textbf{D}(\textbf{r})$ is the diffusivity tensor, a symmetric positive-definite matrix, and $f(\textbf{r})$ is the potential of mean force (PMF) and is related to $\psi(\textbf{r})$ the density profile of the corresponding component  by $\psi(\textbf{r})=\exp[-\beta f(\textbf{r})]$. For a closed uniform system $\psi(\textbf{r})$ is constant and $D_{ij}(\textbf{r})\equiv D\delta_{ij}$ is isotropic. Eq.~(\ref{eq:smoluchowski}) can therefore be solved analytically to obtain~\cite{RevModPhys.15.1}:
\begin{eqnarray}
p(\textbf{r},t|\textbf{r}_0,t_0) &=& \frac{1}{\left[4\pi D(t-t_0)\right]^{d/2}}\exp\left[-\frac{||\textbf{r}-\textbf{r}_0||^2}{4D(t-t_0)}\right]\notag\\ &&\label{eq:smoluchowski_solution_bulk}
\end{eqnarray}
where $d$ is the dimensionality of the system. Ensemble average of $||\textbf{r}(t)-\textbf{r}(0)||^2$ can therefore be calculated from Eq.~(\ref{eq:smoluchowski_solution_bulk}) to obtain the widely known \emph{Einstein formula}:
\begin{eqnarray}
 D &=& \lim_{t\rightarrow\infty}\frac{\left\langle||\textbf{r}(t)-\textbf{r}(0)||^2\right\rangle}{2dt}\label{eq:msd_bulk}
\end{eqnarray}
Another dynamical quantity that can be obtained from $p(\textbf{r},t|\textbf{r}_0,t_0)$ is the structural relaxation time that is calculated from the decay of the self-intermediate scattering function:
\begin{eqnarray}
F_s(q,t) &=& \left\langle 
\frac1N\sum_{i=1}^N\exp[iq|\textbf{r}(t)-\textbf{r}_0|]
\right\rangle
\end{eqnarray}
which is the Fourier transform of $p(\textbf{r},t|\textbf{0},0)$. The decay characteristics of $F_s(q,t)$ is a strong function of temperature and closely mirrors the convergence of mean-square displacement to the diffusive regime. Also, the exact rate of decay depends on $q$ and is typically largest for $q=q_{\max}$, the wavevector corresponding to the first peak of the structure factor $S(q)$. Structural relaxation times are typically obtained by finding the root of $F_s(q_{\max},\tau)=c$ with $c$, being a constant that should be chosen so that it falls into the alpha relaxation region. For three-dimensional systems $c$ is generally chosen to be $1/e$. 

In confined systems that are anisotropic along the $z$ direction however, $\textbf{D}(z)$ is no longer isotropic and has two distinct elements instead: $D^{\parallel}(z)=D_{xx}(z)=D_{yy}(z)$ or the \emph{lateral diffusivity} and $D^{\perp}(z)=D_{zz}(z)$ or the \emph{normal diffusivity}. The $z$ dependence of both these diffusivities turns Eq.~(\ref{eq:smoluchowski_solution_bulk}) into a coupled partial differential equation that cannot be solved via conventional techniques such as separation of variables even though $\psi(z)$ can be accurately determined from a molecular simulation. Several closed-form analytical solutions of Eq.~(\ref{eq:smoluchowski_solution_bulk}) have been reported by making certain \emph{a priori} assumptions about the mathematical form of $\textbf{D}(z)$~\cite{LubenskyPRE2007}. But since $D_{\parallel}(z)$ and $D_{\perp}(z)$ are not \emph{a priori} known, it is only possible to obtain a numerical solution that relates position-dependent diffusivities and the observed $p(\textbf{r}^{\parallel},z,t|\textbf{0},z_0,t_0)$ in a self-consistent manner. In order to avoid this tedious path, different authors have taken different phenomenological approaches for calculating $D^{\parallel}$ profiles. In all of these approaches, the simulation domain is divided into slices, and an \emph{ad hoc} mean-squared displacement, $\phi(z,t)$, is defined for every slice, which, alongside Eq.~(\ref{eq:msd_bulk}), is used to determine lateral diffusivities in each slice. 

A common phenomenological approach used by many authors~\cite{BerneJPhysChemB2004,ShiJCP2011} is to apply the virtual absorbing boundary conditions at the boundaries of each slice. In this approach, the entities that leave and re-enter a slice during a time window do not contribute to the mean-square displacement of that time window since they might spend time in slices with diffusivities different from that of the slice of interest. Therefore $\phi(z,t)$ is defined as:
\begin{eqnarray}
\phi(z,t) &=& \left\langle|\textbf{r}^{\parallel}(t)-\textbf{r}^{\parallel}(0)|^2 \right\rangle_{|z(\tau)-z|<\frac{\Delta z}{2},0\le\tau\le t}\label{eq:msd_adsorbing}
\end{eqnarray}
with $\Delta{z}$ being the thickness of the slice.  This approach is numerically costly since the fraction of particles that do not leave their initial slice during a time interval $t$ will decrease with $t$. As a result, it is far more difficult to gather quality statistics in the long-term diffusive regime of the mean-squared displacement. Another problem with this approach is the sensitivity of the computed relaxation times and diffusivities to the slice thickness due to unacceptable biasing in favor of less mobile particles that are more likely to remain inside a thinner slice. Considering these limitations, some authors have used alternative phenomenological approaches. One possibility is to define $\phi(z,t)$ based on the position of the particle at the \emph{beginning} of the time window~\cite{TeboulJPhysCondMat2002, KumarJCP2005}:
\begin{eqnarray}
\phi(z,t) &=& \left\langle|\textbf{r}^{\parallel}(t)-\textbf{r}^{\parallel}(0)|^2\delta[z(0)-z]\ \right\rangle
\label{eq:msd_start}
\end{eqnarray}

A second possibility is to include in $\phi(z,t)$ only the particles that are present in the slice both in the \emph{beginning} and at the \emph{end} of the time window~\cite{OstrowskyPhysicaA2002}:
\begin{eqnarray}
\phi(z,t) &=& \left\langle|\textbf{r}^{\parallel}(t)-\textbf{r}^{\parallel}(0)|^2\mathfrak{D}(z(0),z(t),z)\ \right\rangle\label{eq:msd_start_end}
\end{eqnarray}
And finally, a third possibility is to assign the contribution of every particle to the $\phi(z,t)$ that corresponds to its average normal position $\bar{z}(t)=(1/t)\int_0^tz(\tau)d\tau$ throughout the time window $t$~\cite{BerendsenJPhysChem1994}:
\begin{eqnarray}
\phi(z,t) &=& \left\langle|\textbf{r}^{\parallel}(t)-\textbf{r}^{\parallel}(0)|^2\delta[\bar{z}(t)-z]\ \right\rangle\label{eq:msd_avg}
\end{eqnarray}
Although arguments can be raised for and against any of these approaches, there is no solid theoretical reason for preferring any one of them over others. In this work, we adopt the approach expressed in Eq.~(\ref{eq:msd_start_end}) due to its computational simplicity and its robustness to changes in the slice thickness. Also, by only allowing the particles that are present in the same slice both at the beginning and at the end of a time window to contribute to $\phi(z,t)$, we avoid the unphysical biasing towards particles that persistently move away from their initial slices towards regions of different diffusivities. 

\begin{figure}
	\begin{center}
		\includegraphics[width=.5\textwidth]{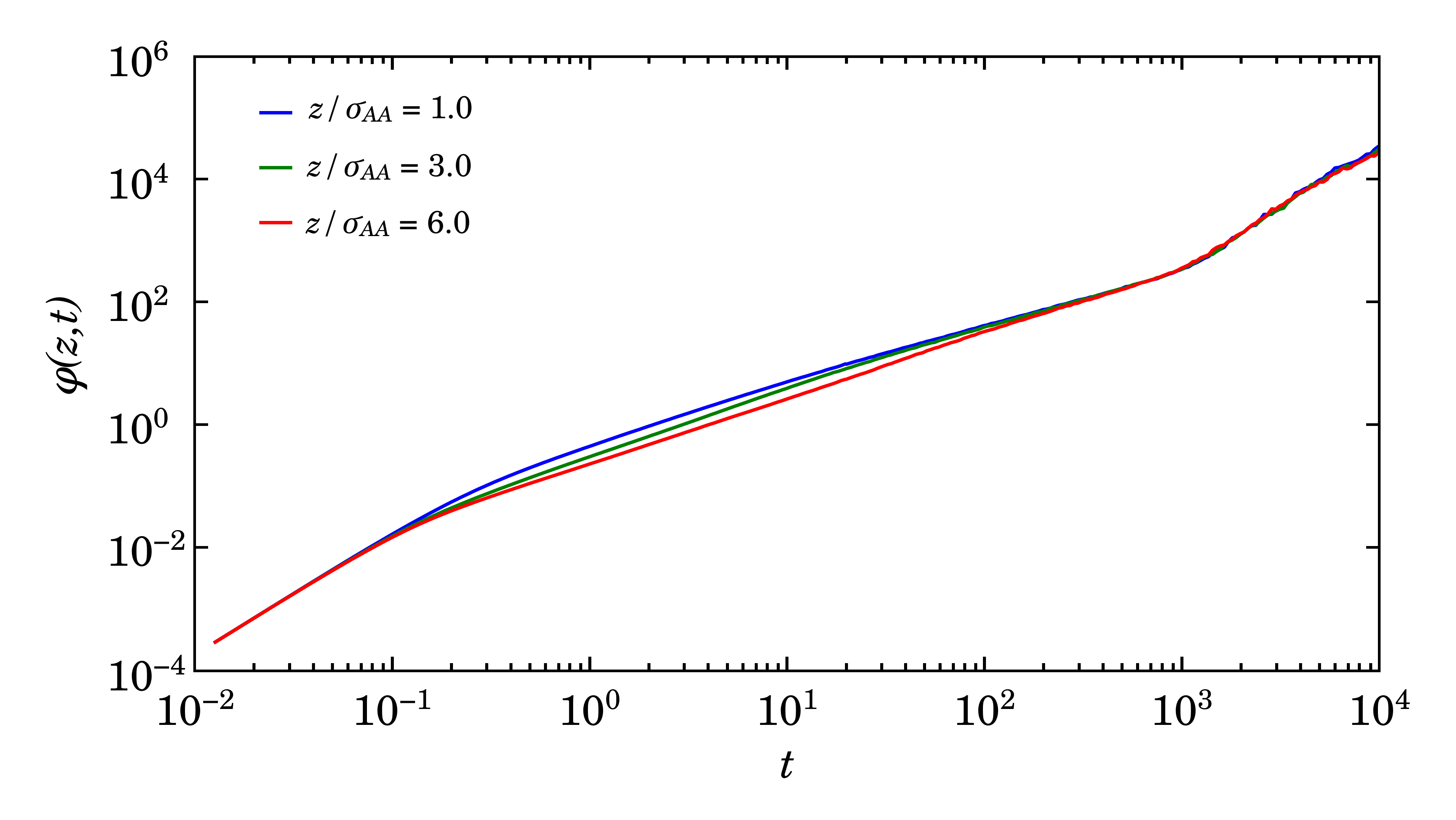}
		\caption{\label{fig:MSD}Total mean-squared displacements computed for a film with $\epsilon_S=0.3$ at $T^*=0.9$. The super diffusive regime at $t\approx10^3$ is due to the inter-slice mixing of particles.
		}
	\end{center}
\end{figure}

Another difficulty in calculating position-dependent diffusivities is the inter-slice mixing that occurs at very large time windows. This leads to an apparent super-diffusive regime in the ad hoc $\phi(z,t)$ at large values of $t$ since at larger time windows, the particles are more likely to take a round-trip to slices that are far away from their slice of origin (Fig.~\ref{fig:MSD}). As a consequence, $D^{\parallel}(z)$'s cannot be computed from the asymptotic behavior of $\phi(z,t)$ and should instead be estimated from the diffusive regime that precedes the mixing-initiated super-diffusive regime. In Fig.~\ref{fig:MSD} for instance, this will correspond to $10^0\le t\le 10^2$.

Using the same convention introduced above, a similar slice scattering function can be defined as follows:
\begin{eqnarray}
F_s(q,z,t) &=& \frac1N\left\langle 
\sum\limits_{i=1}^N
\exp\left\{iq|\Delta\textbf{r}^{\parallel}(t)|\right\}\mathfrak{D}(z(t),z(0),z)
\right\rangle\notag\\&&\label{eq:SISF_conf}
\end{eqnarray}
As previously observed in molecular simulations of two-dimensional systems~\cite{PereraJCP1999}, self-intermediate scattering functions can take lower values during the caging regime in two-dimensional systems than in three-dimensional systems. Therefore the cutoff value for defining relaxation times needs to be smaller in two-dimensional systems, including for $F_s(q,z,t)$'s calculated in this work. We will therefore use a value of $c=0.2$, which always falls in the alpha relation region for all scattering functions calculated in this work. 

\begin{figure}
\begin{center}
	\includegraphics[width=.5\textwidth]{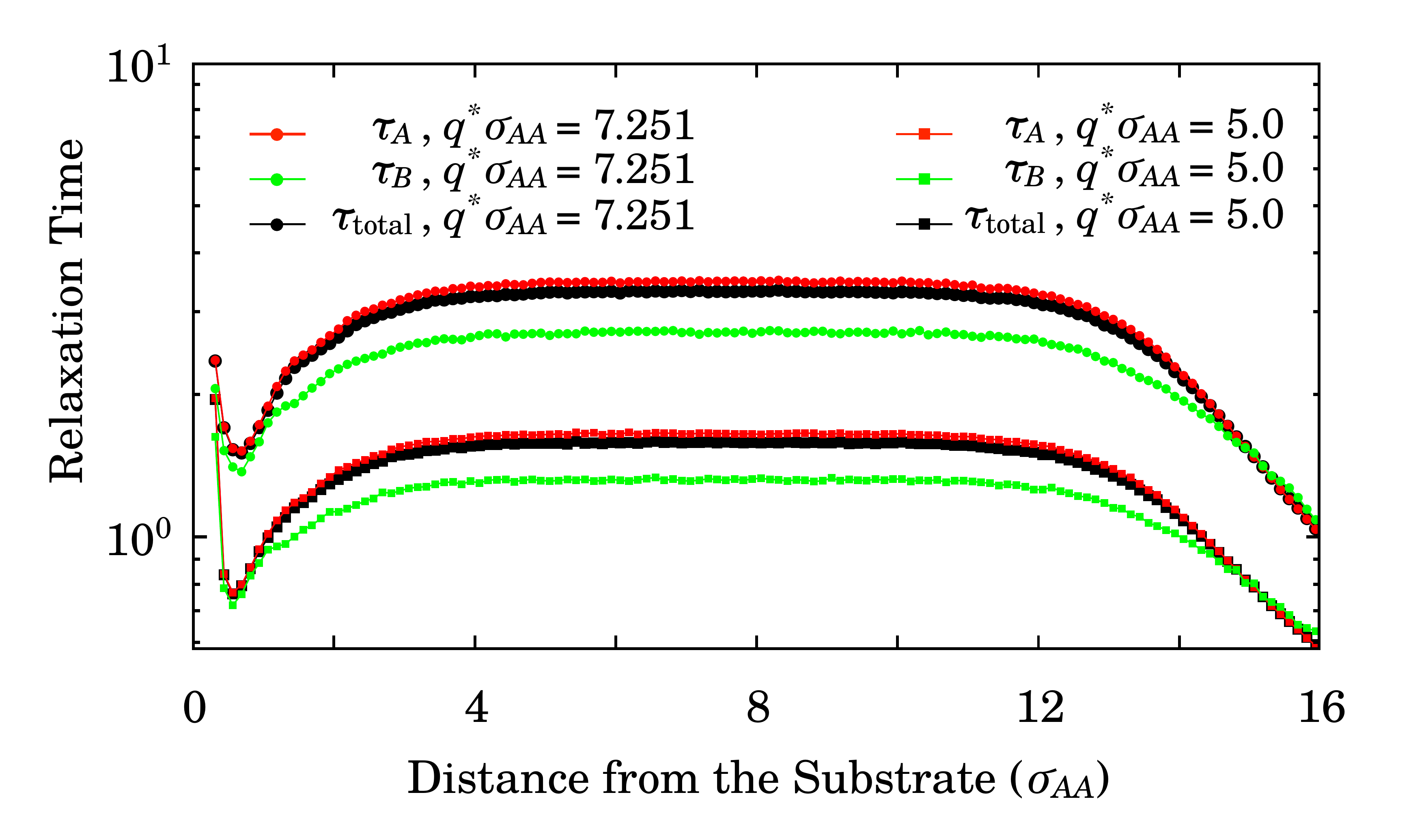}
	\caption{The sensitivity of $\tau^{\parallel}(z)$ to the particular choice of $q$. The lateral relaxation times are computed at two values of $q$ for a film with $\epsilon_S=0.3$ and $T^*=0.9$. Although the precise value of $\tau^{\parallel}(z)$ depends on $q$, the observed trend is independent of the particular choice of $q$.
  \label{fig:RxVsQ}
	}
\end{center}
\end{figure}

\begin{figure*}
	\begin{center}
		\includegraphics[width=.9\textwidth]{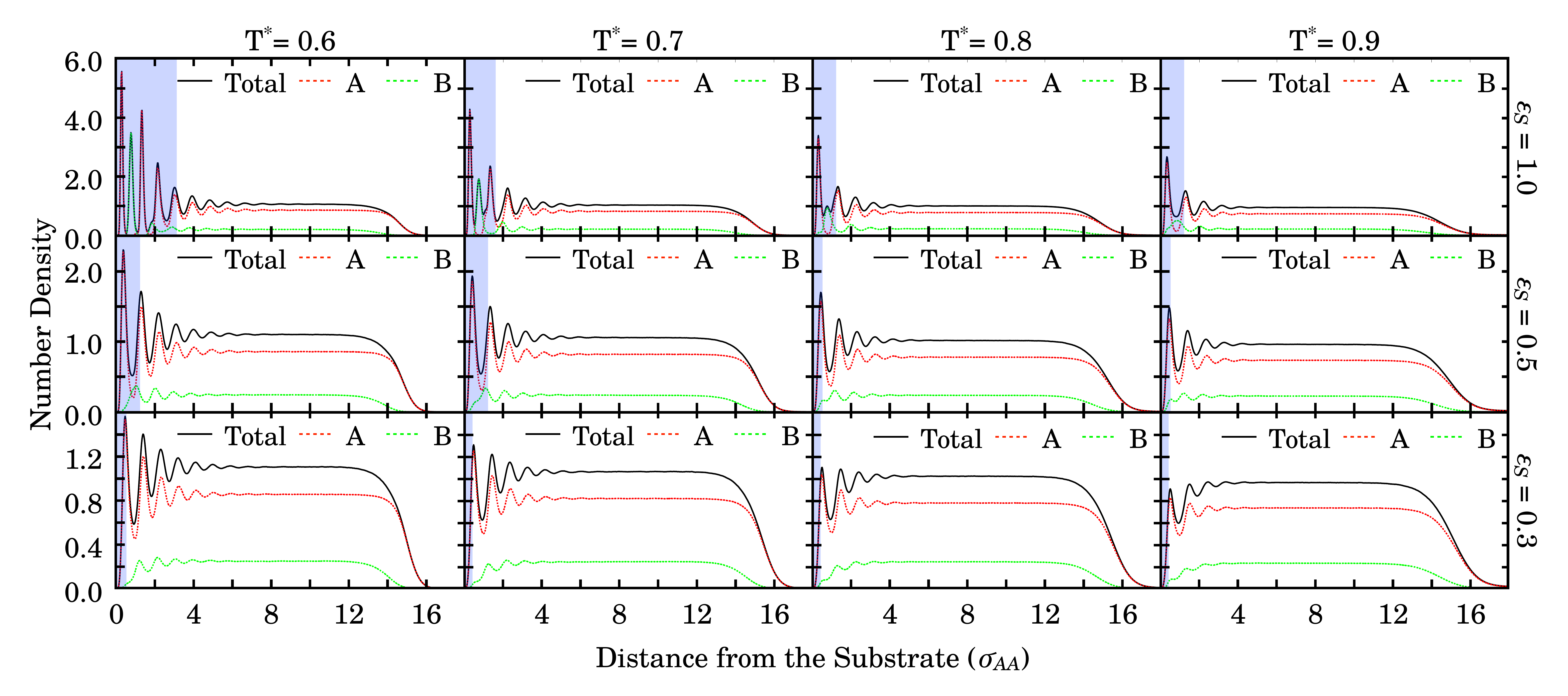}
		\caption{\label{fig:density}Density profiles of thin films as a function of distance from the substrate. The crystalline regions of the films are shaded in light blue. 		
		}
	\end{center}
\end{figure*}

Another ambiguity that arises in computing structural relaxation times in confined systems is the sensitivity of the calculated relaxation time profiles to the particular value of $q$ as the rate of decay in $F_s(q,z,t)$ depends on $q$. Since significant structural variations are possible in a confined system e.g.~due to density and composition oscillations, the position of the first peak of $S(q,z)$ can be a strong function of $z$. However, using different wavenumbers for different regions of space can make comparing the computed relaxation times nontrivial. As a matter of consistency, we use a single $q^*$ value ($q^*=7.251\sigma_{AA}^{-1}$) for all our calculations, and we confirm the robustness of computed relaxation time profiles by computing lateral diffusivities that are independent of the choice of $q^*$. We also confirm that relaxation time profiles are qualitatively insensitive to the particular choice of $q^*$ and the observed trends are preserved if different values of $q^*$ are used even though the exact numerical values of $\tau^{\parallel}$ will change (Fig.~\ref{fig:RxVsQ}).

Due to the constraint imposed in Eqs.~(\ref{eq:msd_start_end}) and (\ref{eq:SISF_conf}), only a fraction of trajectories contribute to $\phi(z,t)$ and $F_s(q,z,t)$ calculations. As a result, it is more difficult to obtain quality statistics in computing these autocorrelation functions. We therefore use thicker slices ($\Delta{z}=\sigma_{AA}/4$) for computing profiles of dynamical quantities.

\section{Results And Discussion\label{results}}
\subsection{Thermodynamic Properties\label{results:static}}

\begin{figure}
    \begin{center}
       \includegraphics[width=.45\textwidth]{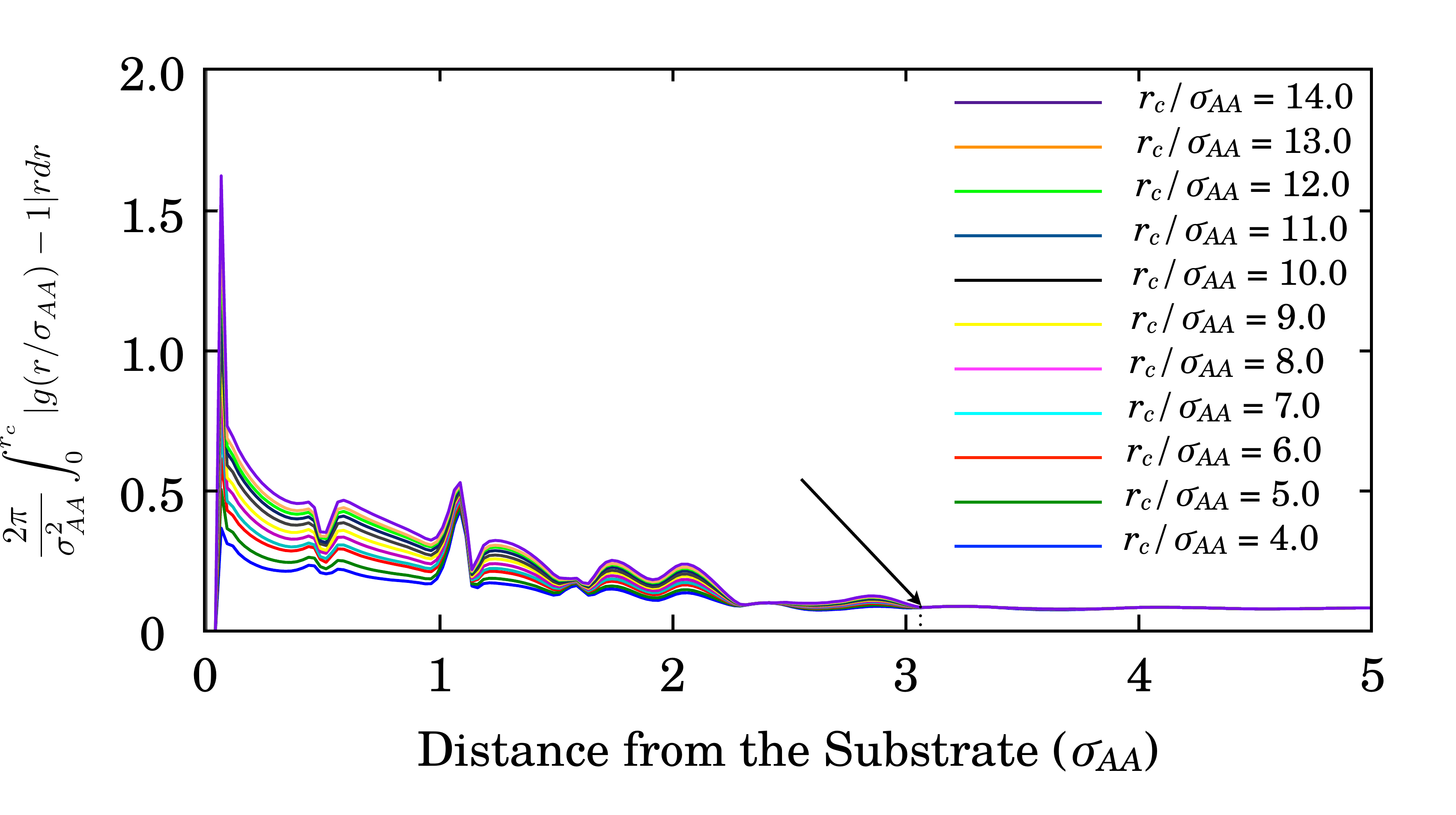}
       \caption{\label{fig:det_xtalinity}Determining the thickness of the crystalline region in the vicinity of the substrate for $T^*=0.6$ and $\epsilon_S=1.0$. The arrow corresponds to the smallest $z$ for which $f(r_c,z;0.05)$ becomes independent of $r_c$.}
    \end{center}
\end{figure}

Figs.~\ref{fig:density} depicts density profiles computed for several values of $T^*$ and $\epsilon_S$. We observe that the liquid is always stratified in the vicinity of the substrate, although the extent of stratification, i.e~the number and the amplitude of density waves, increases by increasing $\epsilon_S$ or by decreasing the temperature. The oscillations in density are both observed for the total density profile (shown in black) as well as the density profiles of individual components (shown in red and green). The emergence of layering in the vicinity of a substrate has been previously observed in a wide variety of other confined systems such as slit pores~\cite{DavisJCP1992} and molecular thin films~\cite{StrioloJPCC2012} and can be attributed to non-random ordering of mobile atoms of the liquid in the close vicinity of immobile atoms of the substrate. These density waves can therefore be conceptually thought of as (unnormalized) substrate-liquid radial distribution functions that are frozen due to the freezing of the atoms in the substrate.

\begin{figure*}
	\begin{center}
		\includegraphics[width=.9\textwidth]{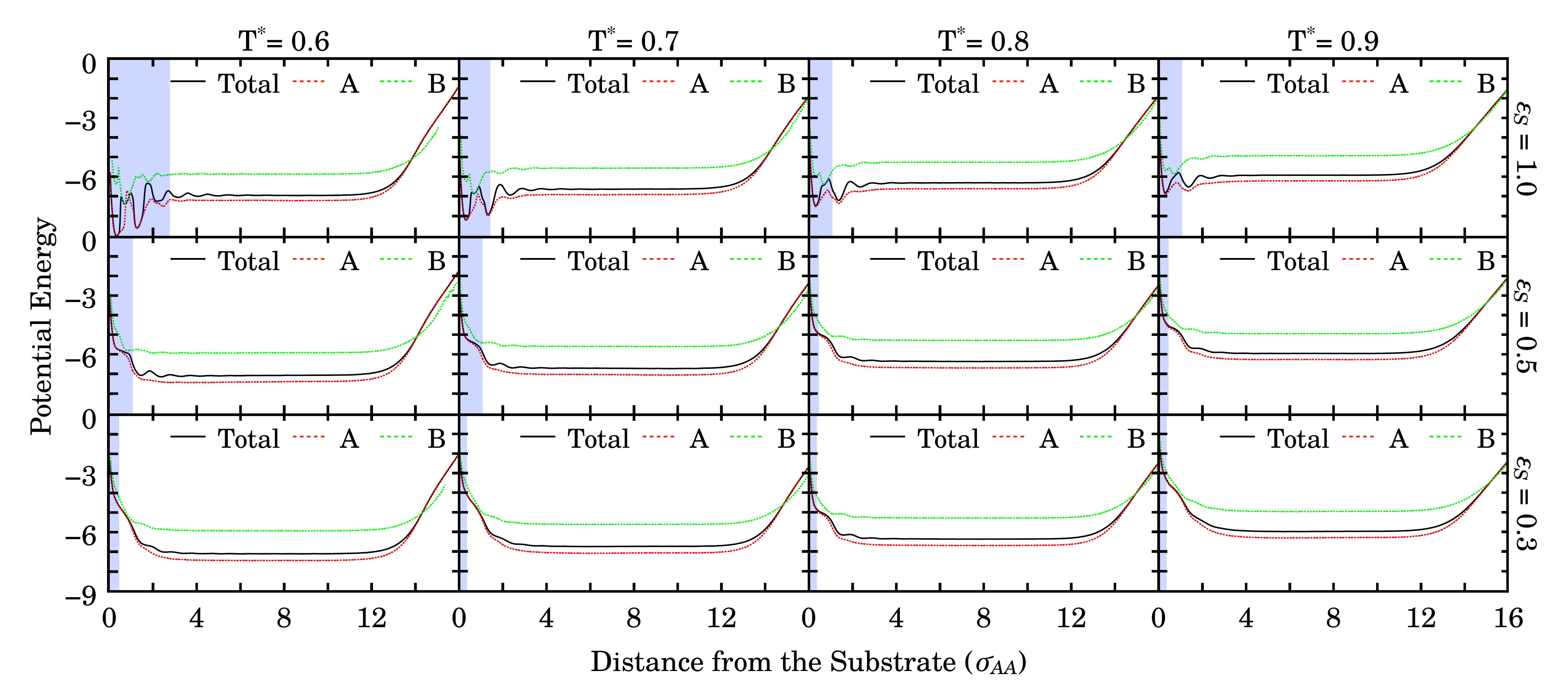}
		\caption{\label{fig:potentialEnergy}Potential energy profiles of thin films as a function of distance from the substrate. 
		The crystalline regions of the films are shaded in light blue.
		}
	\end{center}
\end{figure*}

\begin{figure*}
	\begin{center}
		\includegraphics[width=.9\textwidth]{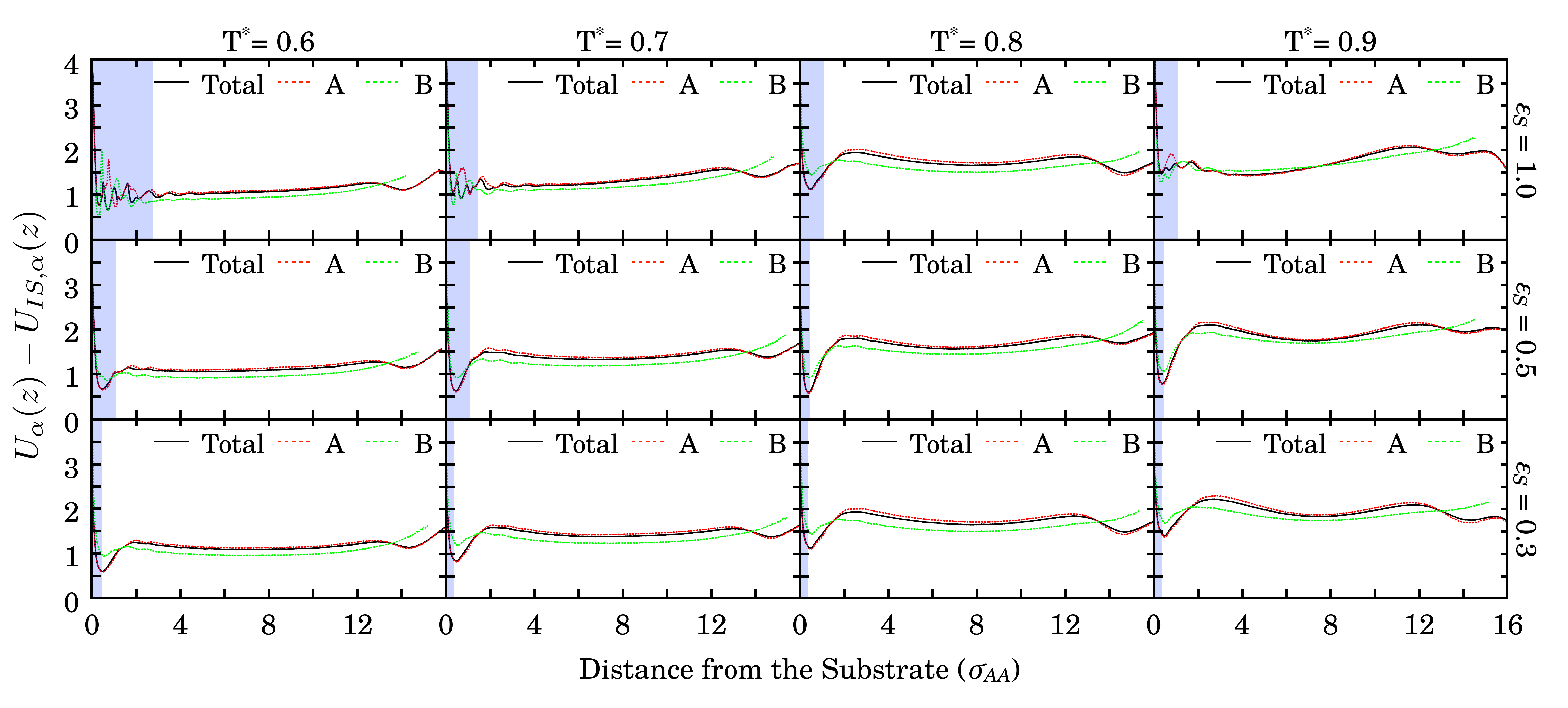}
		\caption{\label{fig:dive}Profile of $\langle U\rangle-\langle U\rangle_{IS}$ of thin films as a function of distance from the substrate. 
	The crystalline regions of the films are shaded in light blue.
	}
	\end{center}
\end{figure*}

The oscillatory density profile of the liquid near the substrate might or might not be accompanied by crystalline order. These two can be distinguished by inspecting the lateral radial distributions computed from Eq.~(\ref{eq:rdf_slice}). In a slice that is laterally amorphous, $g_{\alpha\beta}(r,z)$ converges to unity as $r\rightarrow\infty$ with decaying fluctuations, while in a slice with lateral crystalline order, the fluctuations of $g_{\alpha\beta}(r,z)$ persist over arbitrarily long distances. In order to quantify these two distinct behaviors, we compute the following function for every slice:
\begin{eqnarray}
f_{\alpha\beta}(r_c,z;\nu) &=& \int_0^{r_c}H\big(\big|g_{\alpha\beta}(r,z)-1\big|-\nu\big)dr
\end{eqnarray}
with $H(x)=\int_{-\infty}^x\delta(\xi)d\xi$, the Heaviside function. For amorphous slices, $f_{\alpha\beta}(r_c,z;\nu)$ will be independent of $r_c$ for sufficiently large values of $r_c$ since the amplitude of deviations from unity will be smaller than $\nu>0$ when $r$ is large enough. For crystalline slices, however, $f_{\alpha\beta}(r_c,z;\nu)$ will grow with $r_c$ indefinitely since the characteristic amplitude of deviations from unity will be always larger than a sufficiently small $\nu>0$. We thus define the crystalline region as the part of the film where $f(r_c,z;\nu)$ changes with $r_c$. This procedure is depicted in Fig.~\ref{fig:det_xtalinity}, where $f(r_c,z;0.05)$ vs.~$z$ is given for a few different values of $r_c$ in the simulation conducted at $T^*=0.6$ and $\epsilon_S=1.0$. Since there is no noticeable dependence of $f(r_c,z;\nu)$ on $r_c$ for $z\ge\zeta=3.2\sigma_{AA}$, we define the region $z/\sigma_{AA}\in[0,3.2]$ as the crystalline region of the film. We use the same procedure for determining the widths of crystalline regions ($\zeta$) at other values of $T^*$ and $\epsilon_S$, and show those regions in shaded light blue in Figs.~\ref{fig:density},~\ref{fig:potentialEnergy},~\ref{fig:dive},~\ref{fig:relaxationTime} and~\ref{fig:diffusivity}. Similar values are obtained for $\zeta$ if $g_{AA}(r,z)$ or $g_{BB}(r,z)$ are used instead of $g_{\text{total}}(r,z)$. As can be vividly seen in Fig.~\ref{fig:density}, the width of the crystalline region is always significantly smaller than the width of the stratified region, which shows that crystallinity is not a necessary condition for layering in the vicinity of a substrate. Also the width of the crystalline region increases by decreasing the temperature or by increasing $\epsilon_S$.

\begin{figure}
	\begin{center}
		\includegraphics[width=.48\textwidth]{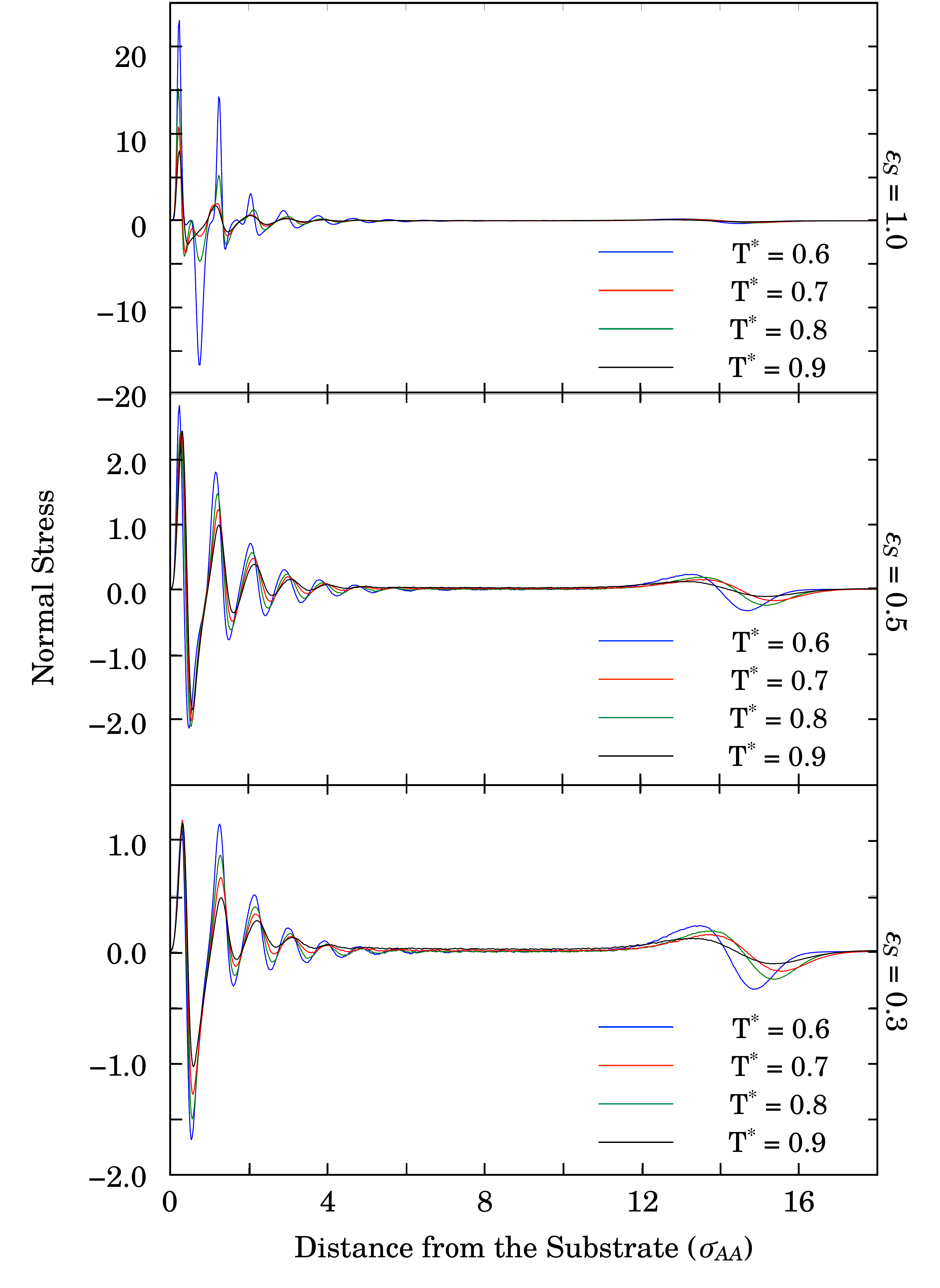}
		\caption{Normal stress profiles for selected values of $\epsilon_S$ and temperature.\label{fig:normalStress}}
	\end{center}
\end{figure}

\begin{figure*}
	\begin{center}
		\includegraphics[width=.85\textwidth]{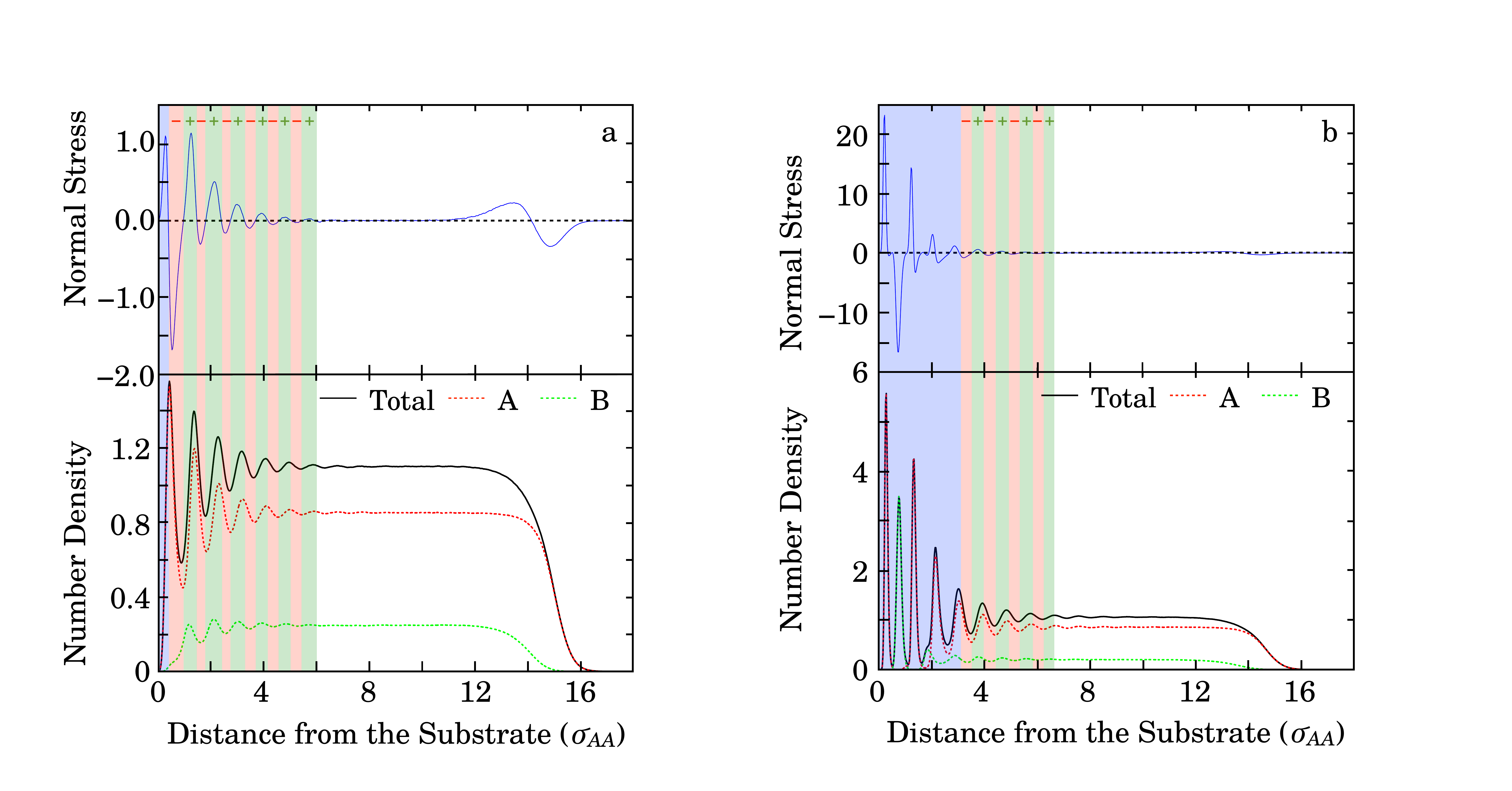}
		\caption{Density and normal stress profiles for (a)~$\epsilon_S=0.3, T^*=0.6$ and (b)~$\epsilon_S=1.0, T^*=0.6$. Regions with lateral long-range order are shaded in blue. In amorphous parts of the film, regions with positive and negative normal stress are shaded in
		 green and red respectively.
		\label{fig:densityVsNormalStress}}
	\end{center}
\end{figure*}

In order to understand the origin of these density oscillations, we inspect the normal stress profiles computed from  Eq.~(\ref{eq:stress}) and depicted in Fig.~\ref{fig:normalStress}. The normal stress profiles are oscillatory for all the temperatures and $\epsilon_S$'s studied in this work, with the depth and amplitude of oscillations increasing upon an increase in $\epsilon_S$ or $1/T$.  (As discussed in Section~\ref{results:dynamic} and depicted in Fig.~\ref{fig:lateralstress}, a similar behavior is observed for oscillations in lateral stress.) Stress oscillations can be relatively strong in the crystalline regions of the films, a behavior expected considering the anisotropic nature of crystals. In amorphous regions of the films however, a direct correlation is observed between density and normal stress, with peaks and valleys of $\rho(z)$ closely following the peaks and valleys of $\mathcal{S}^{zz}(z)$ (Fig.~\ref{fig:densityVsNormalStress}). This is consistent with the  behavior of a mechanically stable fluid that becomes denser upon compression. At higher values of normal stress, that can be thought as the pressure in the $z$ direction, we expect an increase in $\rho(z)$, the marginal-- or plane-averaged-- density in the $z$ direction. Note that this correlation does not hold in the crystalline region (e.g.~Fig.~\ref{fig:densityVsNormalStress}b), since crystals react to stress in nontrivial ways.

\begin{figure}
	\begin{center}
		\includegraphics[width=.47\textwidth]{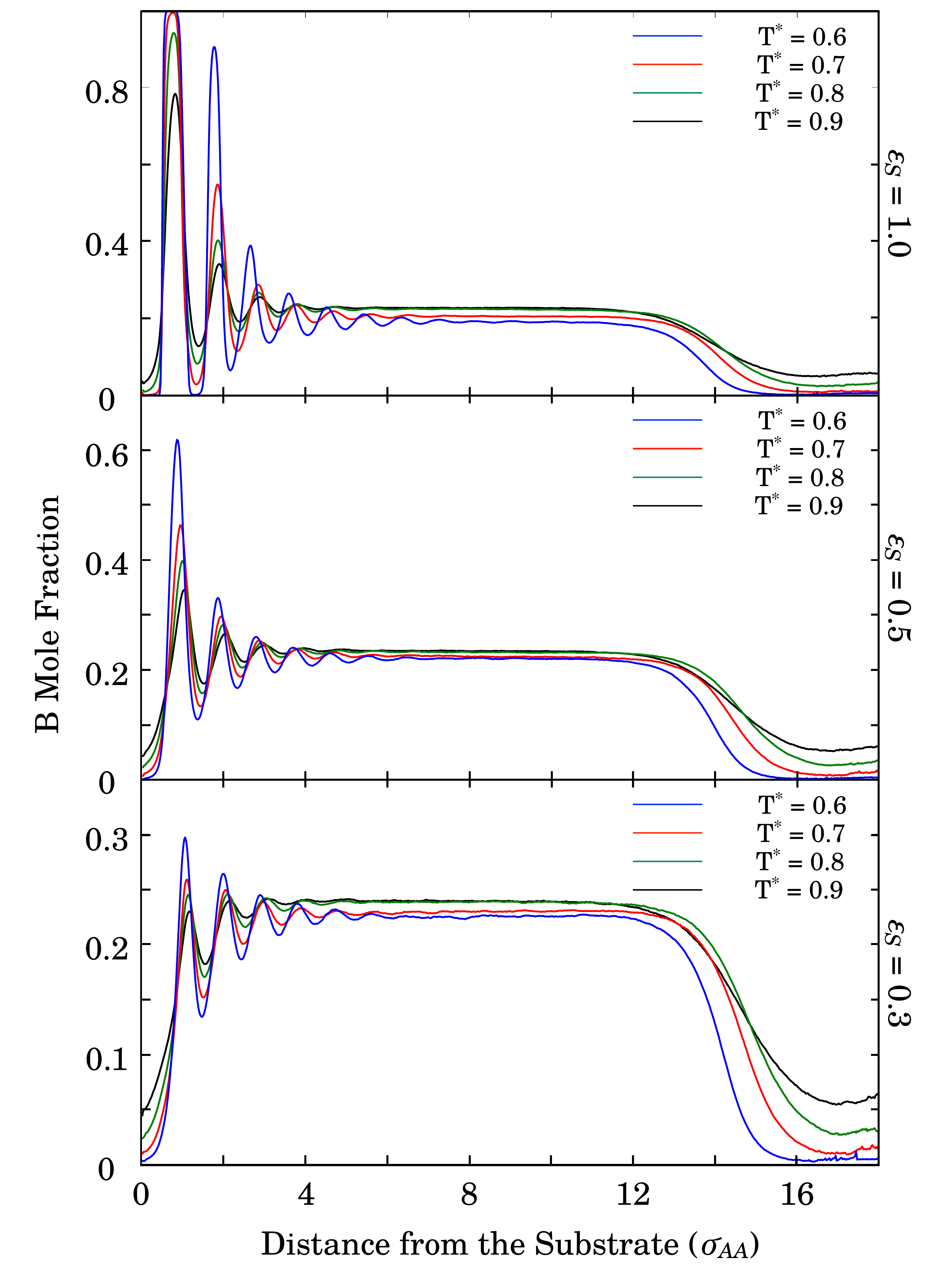}
		\caption{\label{fig:molefraction}The mole fraction of B vs. distance from the substrate for several values of temperature and $\epsilon_S$.}
	\end{center}
\end{figure}

Fig.~\ref{fig:molefraction} depicts the mole fraction of B, $x_B=\rho_B/(\rho_A+\rho_B)$, as a function of distance from the substrate. We observe that the B particles tend to avoid both the liquid-vapor and the liquid solid interfaces. In systems with attractive interactions, the formation-- and the maintenance-- of a free interface would involve an energetic cost arising from the loss of half the nearest neighbor interactions for the particles at the interface. As can be clearly seen in potential energy profiles depicted in Fig.~\ref{fig:potentialEnergy}, the average energy of both particle types increases in the vicinity of the vapor-liquid interface, which is an indication of this energetic loss. 
But this energetic loss is not sufficient for explaining composition differences in multicomponent systems as different composition profiles can give rise to different overall mixing entropies. As shown in detail in Appendix~\ref{appendix:model}, an A-rich interfacial region-- and thus a B-rich bulk region-- will lead to an overall increase in the mixing entropy, which can offset the energetic loss at the interface. 

Due to the presence of attractive A--C and B--C interactions, the energetic cost of maintaining a liquid/substrate interface is smaller, and can be close to non-existent for strongly attractive 'sticky` substrates (e.g.~$\epsilon_S=1.0$). Indeed the average potential energies of both A and B atoms increase in the vicinity of 'loose` substrates (i.e.~substrates with $\epsilon_S=0.3$ and $0.5$), but not to the level of the atoms in the free interface. As a result, we observe an A-rich region in the immediate vicinity of the substrate in all our simulations. However, the mole fraction of B atoms in the interfacial region increases by increasing $\epsilon_S$. This is because the energetic loss is smaller for substrates that are more sticky. 

 Like the profiles of total density, composition profiles are oscillatory near the substrate and $x_B$ undergoes large fluctuations before converging to its bulk value. Also the amplitudes and the number of composition waves are larger at lower temperatures and larger $\epsilon_S$'s. For sticky substrates $(\epsilon_S=1.0)$ and low temperatures $(T^*\le0.7)$, the crystalline region in the vicinity of the substrate is phase-separated into crystalline sheets of A atoms followed by crystalline sheets of B atoms (Top panel of Fig.~\ref{fig:molefraction}).

Fig.~\ref{fig:potentialEnergy} depicts the potential energy profiles calculated from Eq.~(\ref{eq:pot}) for different values of $T^*$ and $\epsilon_S$.  Unlike densities, compositions and lateral and normal stresses that always undergo significant oscillations near the substrate, potential energies tend to increase monotonically in the vicinity of loose substrates, i.e.~for $\epsilon_S=0.3$, and only oscillate near sticky substrates ($\epsilon_S=1.0$). For moderately attracting substrates ($\epsilon_S=0.5$), two distinct regimes are observed at different temperatures. At higher temperatures $(T^*\ge0.8$), potential energy profiles are monotonic and are like the ones observed for $\epsilon_S=0.3$. For lower temperatures, however, they are more oscillatory near the substrate, like the ones observed for sticky substrates. The emergence of oscillatory potential energy profiles can be attributed to the strong ordering induced by these sticky substrates, something that also manifests itself in stronger oscillations in density, stress and composition profiles. This effect is much weaker in the vicinity of looser substrates, giving rise to more monotonic potential energy profiles. 

Fig.~\ref{fig:dive} depicts $U_{\alpha}(z)-U_{IS,\alpha}(z)$ for different values of $T^*$ and $\epsilon_S$. This quantity-- that we call the \emph{dive profile}-- is the average energy that particles of type $\alpha$ residing at $z$, lose as a result of energy minimization, and is a measure of the efficiency with which particles of a certain region can explore the potential energy landscape. We observe larger values of $U_{\alpha}(z)-U_{IS,\alpha}(z)$ in the vicinity of the vapor-liquid interface, which corresponds to more efficient sampling of the potential energy landscape by these particles. This has also been observed in computational studies of free-standing thin films of the Kob-Andersen LJ mixture~\cite{ShiJCP2011}. A similar maximum is observed in the subsurface region neighboring loose substrates. In the vicinity of sticky substrates, however, 
the potential energy landscape is more heavily affected by the presence of the substrate, and the particles are more restricted in sampling and exploring that oscillatory landscape. Similar to potential energy profiles, dive profiles are also very oscillatory in the vicinity of sticky substrates, which is also a consequence of strong substrate-induced ordering in those regions.  Another interesting observation is the gradual mild decline of $U_{\alpha}(z)-U_{IS,\alpha}(z)$ across the films that are in the vicinity of sticky substrates. This shows that the ordering effect of a strongly attractive substrate extends far beyond the solid/liquid subsurface region. This 'deep` ordering induced by sticky walls also manifests itself in stronger stratification in the liquid film.

\subsection{Kinetic Properties\label{results:dynamic}}

\begin{figure*}
	\begin{center}
		\includegraphics[width=.9\textwidth]{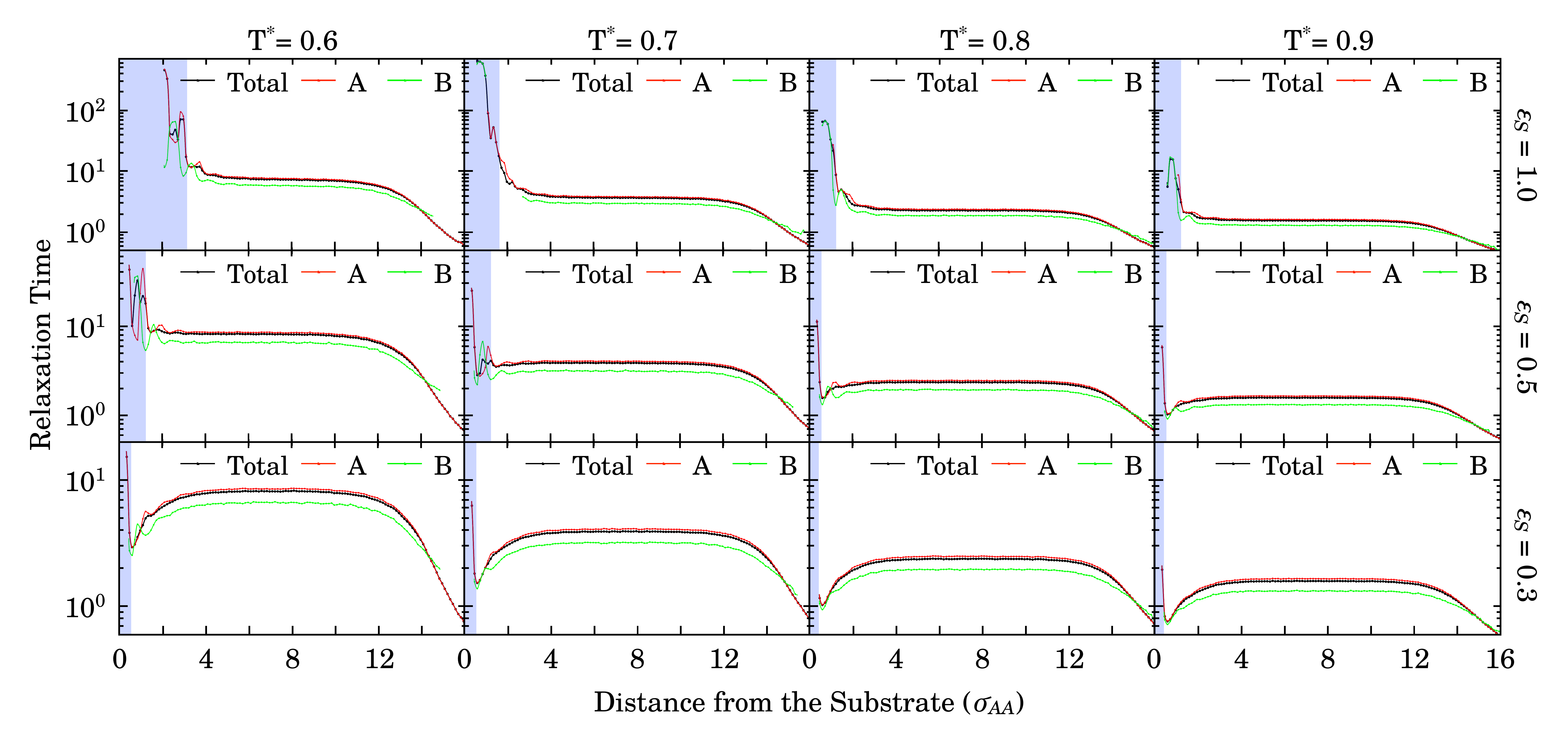}
		\caption{Relaxation time profiles of thin films as a function of distance from the substrate as calculated from the self intermediate scattering functions. The crystalline regions of the films are shaded in light blue.\label{fig:relaxationTime}}
	\end{center}
\end{figure*}

\begin{figure*}
	\begin{center}
		\includegraphics[width=.9\textwidth]{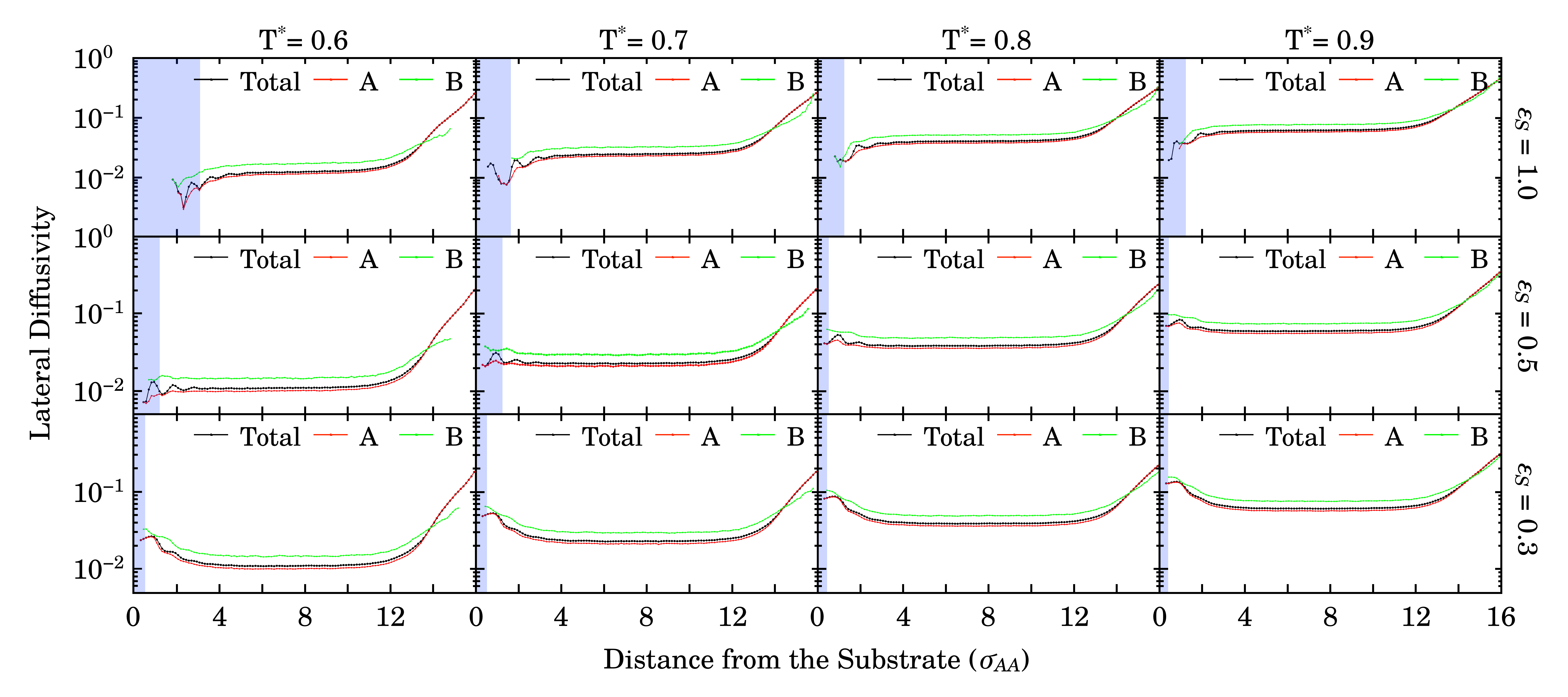}
		\caption{Lateral diffusivity profiles of thin films as a function of distance from the substrate as calculated from the mean squared displacements.The crystalline regions of the films are shaded in light blue.
		\label{fig:diffusivity}}
	\end{center}
\end{figure*}

We compute position-dependent structural relaxation times and lateral diffusivities using the convention introduced in Section~\ref{methods:dynamical}. The results are depicted in Figs.~\ref{fig:relaxationTime} and~\ref{fig:diffusivity} respectively. We observe a region of accelerated dynamics close to the free interface, characterized by large diffusivities and small relaxation times. This is consistent with earlier experimental and computational studies of thin films~\cite{BellJACS2003, ZhuEdigerPRL2011, ShiJCP2011}. 
Also, B atoms typically have larger diffusivities and smaller relaxation times, a fact that can be attributed to their relatively smaller size ($\sigma_{BB}$) that helps them move past other particles more easily than the bulkier A particles. 

The dynamics in the solid-liquid interfacial region is more interesting. In the immediate vicinity of the substrate, dynamics is slower than in the bulk for all values of $T$ and $\epsilon_S$. This is a direct consequence of crystalline order in those regions, and leads to larger relaxation times and smaller diffusivities in the shaded blue regions of Figs.~\ref{fig:relaxationTime} and~\ref{fig:diffusivity}. In highly-ordered crystalline regions of the film, structural relaxations can be too slow to be accessible on the time scales of our simulations. This explains the absence of symbols for some values of $z$ in Figs.~\ref{fig:relaxationTime} and~\ref{fig:diffusivity}, since unlike thermodynamic properties that can be computed with arbitrary accuracy for all values of $z$, dynamical properties cannot be computed for the values of $z$ at which structural relaxation is prohibitively slow.

\begin{figure}
\begin{center}
	\includegraphics[width=.4\textwidth]{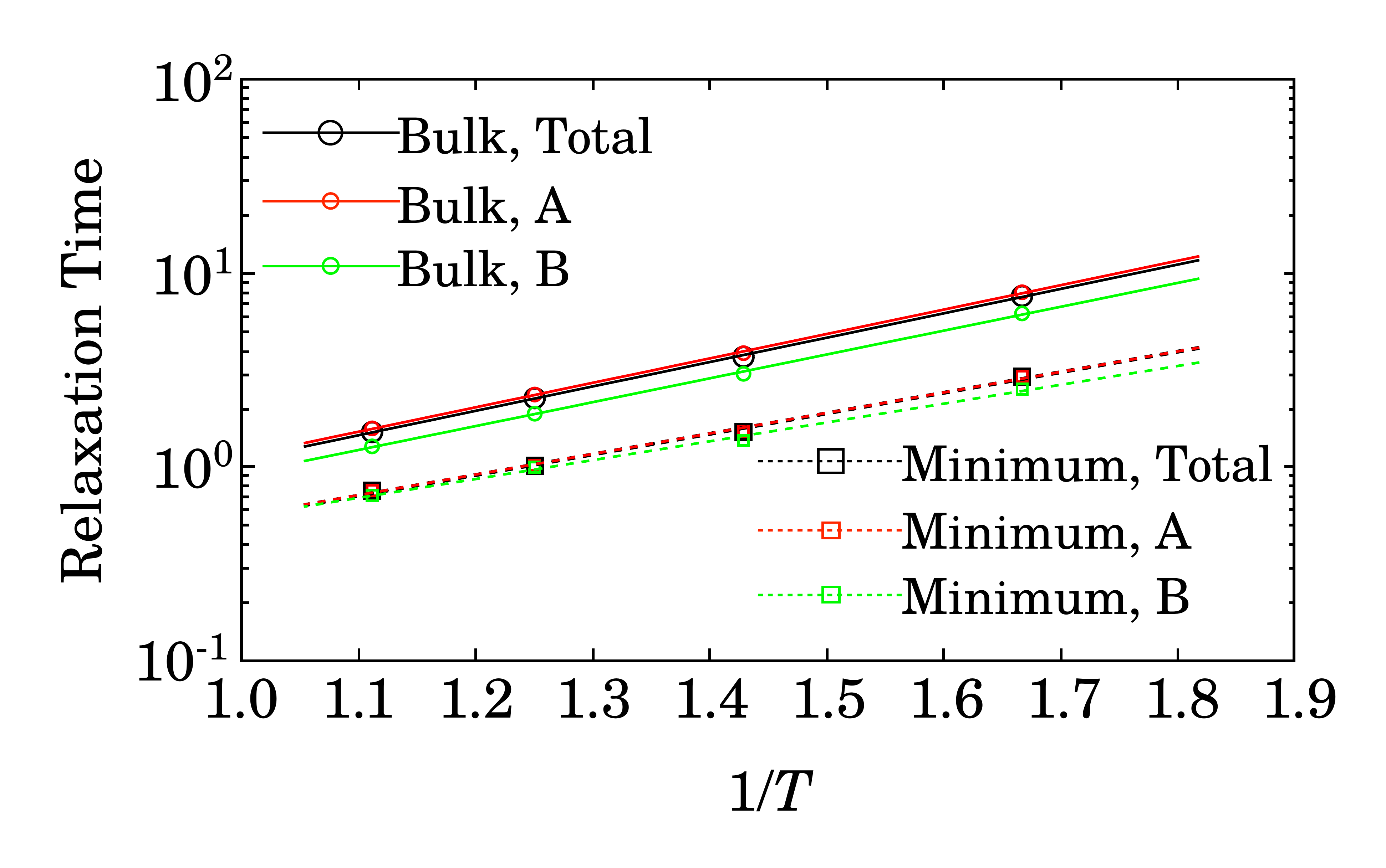}
	\caption{
	$\tau_{\min}^{\parallel}$ and $\tau_{\text{bulk}}^{\parallel}$ as a function of temperature for $\epsilon_S=0.3$. All relaxation times follow the Arrhenius relationship.
	\label{fig:Rlx03Arrhenius}
	}
\end{center}
\end{figure}

\begin{figure}
	\begin{center}
		\includegraphics[width=.45\textwidth]{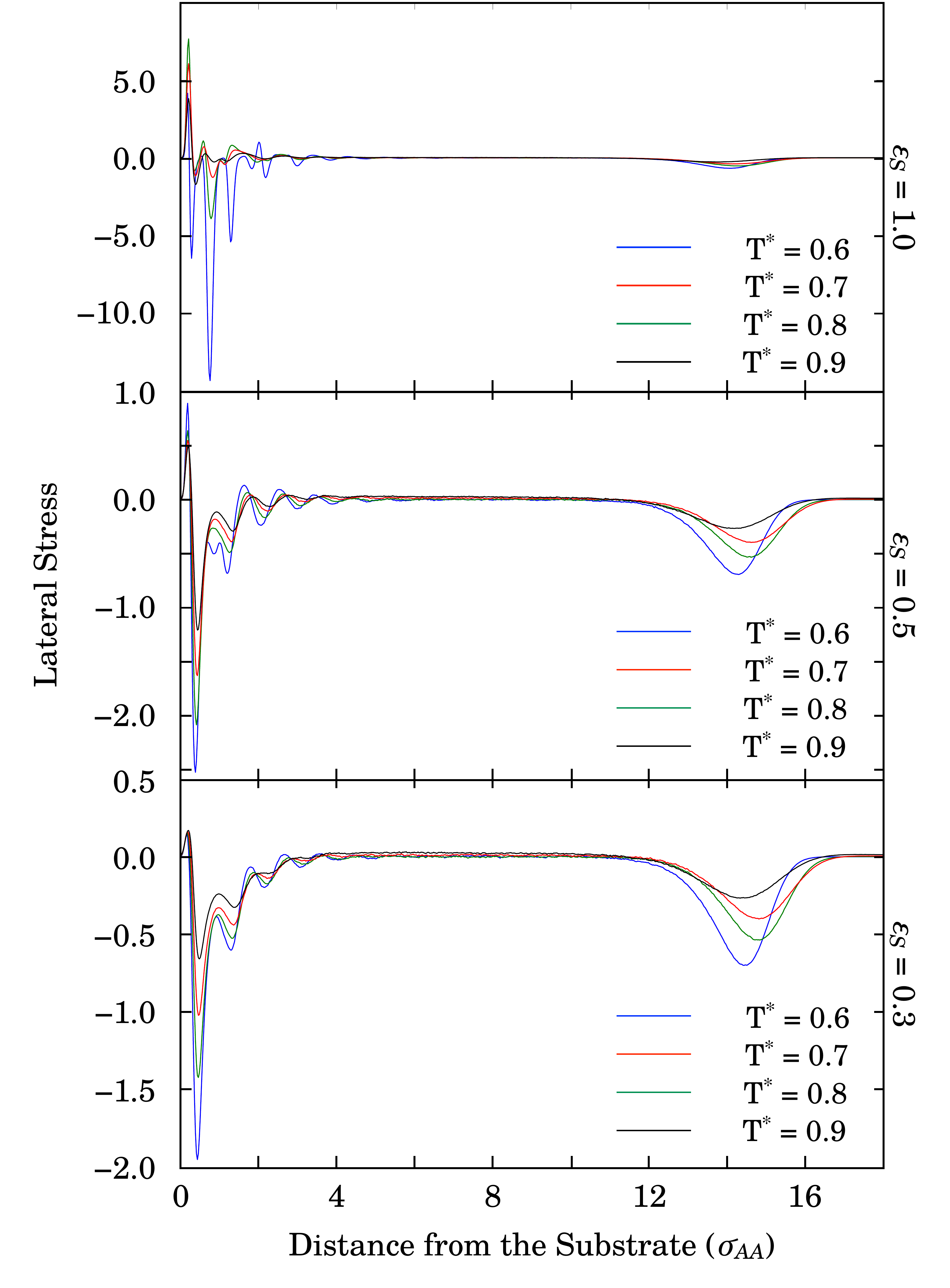}
		\caption{Lateral stress profiles for several values of $\epsilon_S$ and temperature.\label{fig:lateralstress}}
	\end{center}
\end{figure}

\begin{figure*}
\begin{center}
	\includegraphics[width=.8\textwidth]{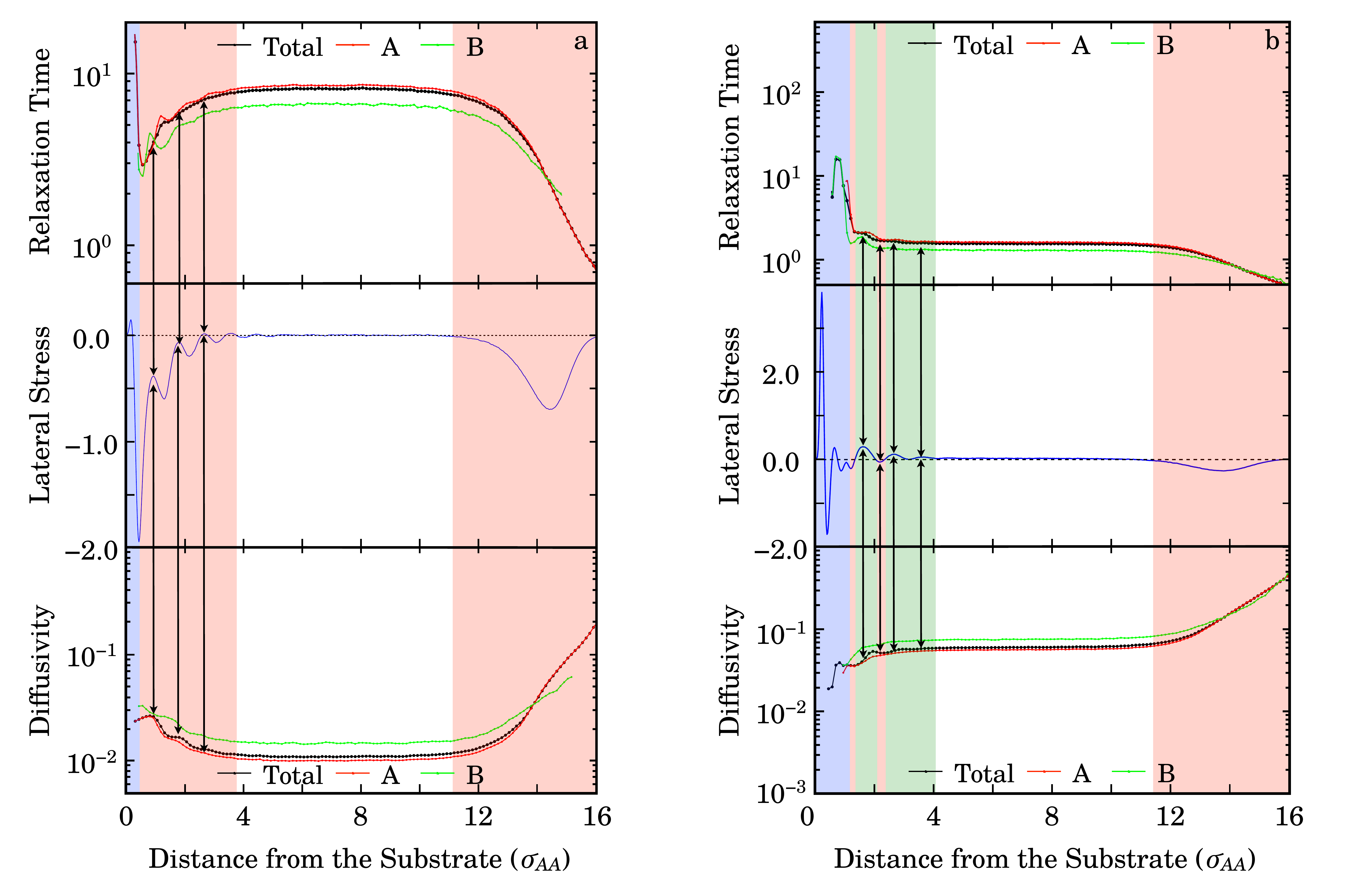}
	\caption{Lateral stress vs. dynamical properties for (a)~$T^*=0.6, \epsilon_S=0.3$ and (b)~$T^*=0.9, \epsilon_S=1.0$. The crystalline regions of the films are shaded in blue, while the amorphous regions with positive and negative lateral stress are shaded in green and red respectively.
	\label{fig:latStressVsDyn}
	}
\end{center}
\end{figure*}

\begin{figure*}
\begin{center}
	\includegraphics[width=.9\textwidth]{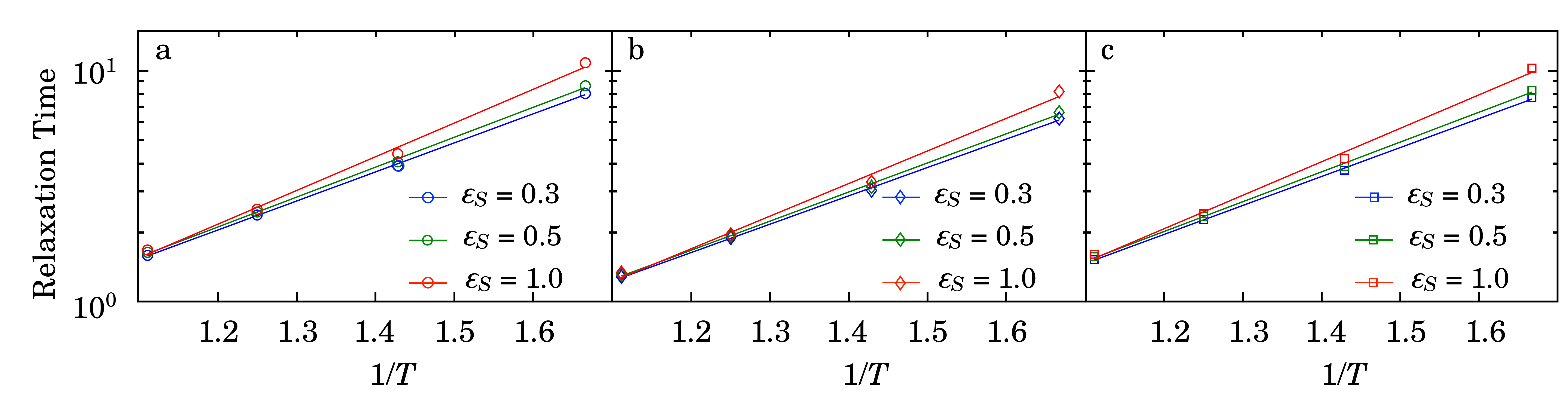}
	\caption{
	Temperature dependence of bulk relaxation times. Panels (a) and (b) correspond to the A ($\circ$) and B ($\diamond$) particles respectively while panel (c) gives the total ($\square$) bulk relaxation times.
	\label{fig:BulkRlxArrhenius}
	}
\end{center}
\end{figure*}

\begin{figure}
	\begin{center}
		\includegraphics[width=.5\textwidth]{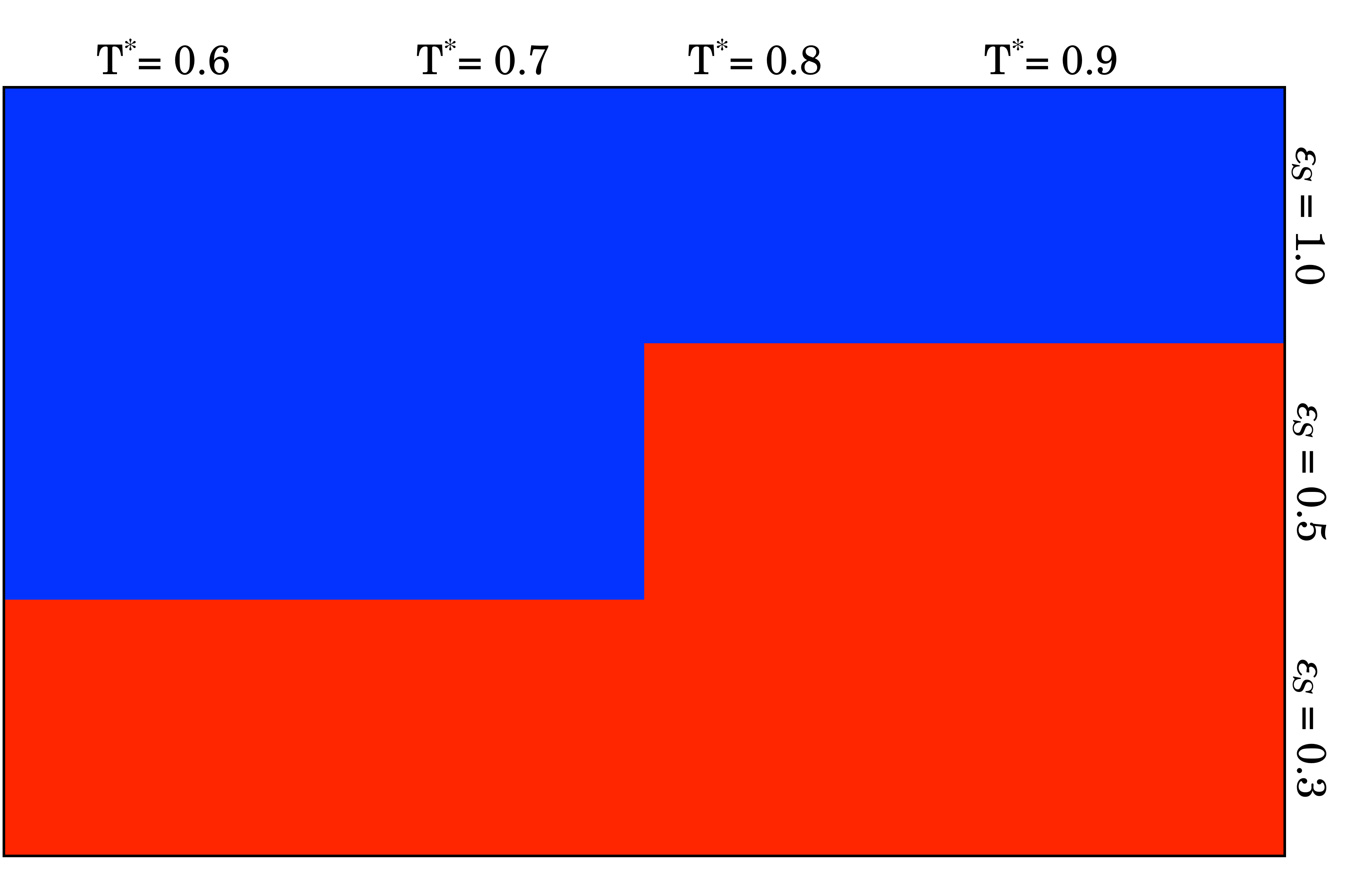}
		\caption{\label{fig:summary}
		The schematic phase diagram of the thin film system. Two distinct regimes are depicted in blue and red respectively.  For the state points depicted in blue, dynamics is decelerated near the substrate, and the potential energy profiles are oscillatory. For the state points depicted in red however, the dynamics is accelerated in the amorphous region near the substrate and the potential energy profile is monotonic. The subsurface region explores the potential energy landscape more efficiently in the state points depicted in red.  
		}
	\end{center}
\end{figure}

\begin{figure}
	\begin{center}
		\includegraphics[width=.5\textwidth]{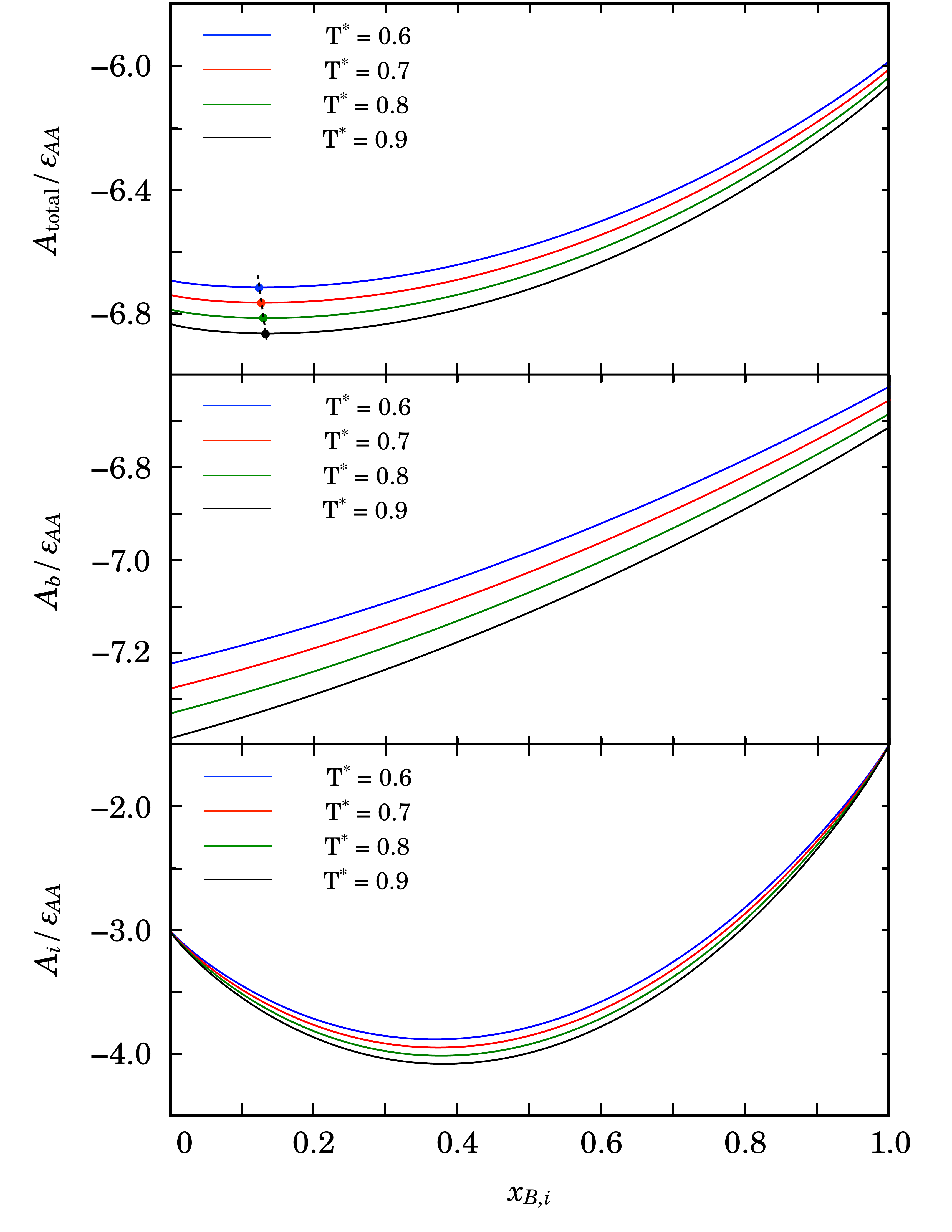}
		\caption{\label{fig:helmapprox}The bulk, interfacial, and total free energy of the two-state model presented in Appndix~\ref{appendix:model}. The dashed line in the top panel corresponds to the loci of global minima of $A_{\text{total}}/\epsilon_{AA}$. }
	\end{center}
\end{figure}

Unlike the crystalline regions that always have slower dynamics than the bulk, we observe two distinct dynamical regimes for the amorphous regions that appear immediately after those crystalline regions. For loose substrates, i.e.~$\epsilon_S=0.3$, the dynamics is accelerated in the amorphous region close to the substrate. This leads to a local minimum for $\tau^{\parallel}(z)$-- and a local maximum for $D^{\parallel}(z)$-- since the dynamics slows down again in the crystalline region of the film. The extent of dynamic acceleration in the vicinity of the substrate is appreciable, and $\tau_{\min}^{\parallel}$, the local minimum of the relaxation time, can be $2-3$ times smaller than $\tau_{\text{bulk}}^{\parallel}$, the bulk relaxation time. 

Fig.~\ref{fig:Rlx03Arrhenius} shows $\tau_{\min}^{\parallel}$ and $\tau_{\text{bulk}}^{\parallel}$ vs. $1/T$, with the bulk relaxation time defined as $\tau^{\parallel}_{\text{bulk}}=(1/5)\int_{5}^{10}\tau^{\parallel}(\bar{z})d\bar{z}$ with $\bar{z}:=z/\sigma_{AA}$. Both $\tau_{\min}^{\parallel}$ and $\tau_{\text{bulk}}^{\parallel}$ have an Arrhenius-type dependence on temperature. Indeed the deviations from  Arrhenius behavior are only significant around the mode coupling temperature, which is $T_c^*=0.435$ for the Kob-Andersen LJ mixture at a reduced number density of $1.20$~\cite{KobAndersenPRE1995}, 
{For the thin films studied here, $T_c$ might be slightly different at different parts of the film due to fluctuations in density and composition. However, such differences are usually very small, and the lowest temperature studied in this work, $T^*=0.6$, will always be much higher than $T_c^*$. The substrate-induced dynamical acceleration becomes stronger as the temperature decreases, which is apparent in the larger slope of the bulk relaxation times in Fig.~\ref{fig:Rlx03Arrhenius}. Also, the dynamical landscape becomes more oscillatory at lower temperatures, with the appearance of further maxima and minima in the relaxation time and diffusivity profiles.

We observe the completely opposite behavior in the vicinity of sticky substrates, i.e.~for $\epsilon_S=1.0$, where both the structural relaxation and diffusion are slower in the amorphous region close to the substrate. This dynamical deceleration becomes more pronounced at lower temperatures, and the rate of structural relaxation can decrease by several orders of magnitude, as can be seen in the top panel of Figs.~\ref{fig:relaxationTime} and~\ref{fig:diffusivity}. Also the dynamical landscape of the film is highly oscillatory, with diffusivity and relaxation time profiles showing several maxima and minima across the film. 

For $\epsilon_S=0.5$, namely a moderately attractive substrate, the qualitative features of the observed dynamical regime depend on temperature. At higher temperatures, the dynamics is accelerated in the amorphous region close to the substrate, similar to the behavior observed for $\epsilon_S=0.3$. However, the magnitude of this dynamical acceleration is smaller. At $T^*=0.9$ for instance, $\tau_{\text{bulk}}^{\parallel}/\tau_{\min}^{\parallel}\approx1.52$ which is smaller than $2.01$, the value observed for $\epsilon_S=0.3$ at the same temperature. Upon decreasing temperature, this dynamical acceleration becomes weaker and weaker, until a cross-over occurs at $T^*\approx0.7$ where a shift to the substrate-induced decelerated dynamics  is observed.

These two distinct dynamical regimes can be explained by inspecting the lateral stress profiles depicted in  Fig.~\ref{fig:lateralstress}. Like the normal stress that is a pressure in the $z$ direction, lateral stress is a pressure in the $xy$ plane. In simple fluids, dynamics become slower at higher pressures. It is therefore natural to expect that higher lateral stress will also lead to slower lateral dynamics. As can be seen in Fig.~\ref{fig:latStressVsDyn}, oscillations of dynamical properties closely follow the oscillations in lateral stress. The dynamical acceleration in the vicinity of weakly attractive substrates can therefore be attributed to tensile lateral stress in those regions, while in the vicinity of strongly attractive substrates, the lateral stress is typically compressive, which leads to a slowdown in dynamics. The temperature dependence of these two dynamical regimes can also be explained by inspecting the lateral stress profiles. As temperature deccreases, the magnitude of tensile or compressive lateral stress becomes larger. As a consequence, the acceleration or deceleration of dynamics (with respect the bulk) becomes stronger as well.

Unlike the oscillations in thermodynamic properties that can extend deep into the bulk region of the film, dynamical properties tend to converge much more quickly to their bulk values. As a result, bulk relaxation times can be readily calculated for all the films studied in this work. For $\epsilon_S=0.3$ and $\epsilon_S=0.5$, relaxation times converge to the bulk value relatively quickly. We therefore define the bulk relaxation time as the average relaxation time for $5\le z/\sigma_{AA}\le10$. For $\epsilon_S=1.0$ however, this convergence is slower and occurs at a further distance from the substrate. We therefore use the range $6\le z/\sigma_{AA}\le 10$ for this calculation. Fig.~\ref{fig:BulkRlxArrhenius} depicts the temperature dependence of $\tau^{\parallel}_{\text{bulk}}$. The temperature dependence is satisfactorily described by the Arrhenius relationship, which is not surprising considering the relatively high temperatures studied in this work.

\section{Conclusions\label{discussion}}
In this work, we find that a substrate can induce large oscillations in a wide range of thermodynamic and dynamical properties across a film.  Properties such as density, composition, and stress can undergo large oscillations near a substrate. We explain the emergence of  density waves by establishing a correlation between density oscillations and normal stress oscillations. This is in line with our intuition that mechanically stable fluids  become denser in the presence of compressive normal stress. We also propose a simple thermodynamic model to explain the preference of B atoms to avoid interfacial regions. Unlike densities, compositions, and stresses, we observe potential energy profiles to oscillate only in the vicinity of sticky substrates. 

We investigate dynamical anisotropies of the film by computing position-dependent diffusivities and relaxation times and discover two distinct dynamical regimes in the vicinity of the substrate. For sticky substrates and/or low temperatures, structural relaxation is decelerated in the solid-liquid interfacial region, a behavior that is in accordance with our intuition that a substrate will lead to dynamical deceleration. For loose substrates and/or high temperatures, however, we observe a counter-intuitive acceleration of dynamics in the amorphous liquid near the substrate. We explain both these regimes by studying lateral stress profiles across the film and establish a strong correlation between the oscillations in lateral stress and the oscillations in the corresponding kinetic properties.

Consequently, two distinct qualitative behaviors are observed for energetics and dynamics in the vicinity of attractive substrates that are summarized in Fig.~\ref{fig:summary}. Looser substrates are characterized by faster dynamics, a monotonic increase in potential energy, and a more efficient exploration of the potential energy landscape in subsurfaces of solid-liquid interfaces. In this context, the behavior of the solid-liquid subsurface is very similar to that of the vapor-liquid subsurface observed in earlier computational studies of free-standing thin films~\cite{ShiJCP2011}. 
Near sticky substrates however, we observe the completely opposite regime, characterized by strong ordering, slower dynamics, and an oscillatory potential energy landscape in the solid-liquid subsurface region.  For moderately interacting substrates, both these regimes are possible depending on temperature.  These findings are very important in the context of vapor-deposited stable glasses discussed in Section~\ref{section:intro}, since it has been argued in the original publication of Swallen~\emph{et al} (Ref.~\cite{EdigerScience2007}) that accelerated dynamics in the vapor-liquid subsurface can lead to more efficient exploration of the potential energy landscape, and the formation of ultrastable glasses. Our findings suggest that similar dynamical acceleration can also occur in a solid-liquid interface. This dynamical acceleration can be employed to develop alternative procedures for making stable glasses in cases when the vapor-deposition process is not practical, e.g.
~due to possible chemical reactions in the gas phase, or for safety reasons. It can also guide experimental efforts in identifying or designing better substrates for the deposition process.

For all the films studied in this work, the computed structural relaxation times are typically higher in the crystalline regions than in the bulk. Nevertheless, there is a difference between the crystals observed in this work and bulk crystals in terms of dynamical behavior since there is always an interfacial region that separates the crystalline region and the bulk liquid. In that boundary region, dynamics is faster than the bulk crystal while being slower than the bulk liquid.

In glass-forming liquids, the temperature dependence of dynamical properties, such as relaxation times and transport coefficients, can deviate significantly from the generic Arrhenius behavior~\cite{DebenedettiNature2001}. This non-Arrhenius behavior has been studied and well documented for the bulk Kob-Andersen LJ system~\cite{SastryLJ-JPCM2004}. We do not observe this fragility in the temperature dependence of bulk relaxation times calculated due to the relatively high temperatures considered in this work, although we expect the fragility to arise in films that have lower temperatures.  In order to understand the effect of a substrate on the fragility of a liquid film, further simulations on a wider range of temperatures are necessary so that more rigorous analysis can be performed using correlations such as the Vogel-Fulcher-Tammann (VFT) law~\cite{RaultVFT1997}. We believe that this could be an interesting topic for future studies.

Thanks to the large system sizes studied in this work, a bulk region develops between the two interfaces. This allows us to study the effect of these two interfaces independently. Therefore, we do not expect our findings to be altered if thicker films are studied. A topic of interest that could be the subject of further studies is to simulate ultra-thin films in which there is an overlap between the two interfacial regions, and to understand how these interfacial regions affect the structural and dynamical features of the corresponding subsurface regions. 

We confine ourselves to atomic films that are in the vicinity of rough substrates, i.e.~substrates made of explicit atoms that are of comparable size to the liquid atoms, and that are arranged in an fcc lattice. Other alternatives that have been used in earlier studies of confined systems are the smooth implicit substrates such as the LJ 9-3 substrate~\cite{SpohrJCP1997} or the LJ 10-4 substrate~\cite{DavisJCP1992}, or rough amorphous substrates~\cite{SinghNatMater2013}. As has been shown in earlier studies  of slit pores, the qualitative behavior of the fluid is not very sensitive to structural details of the substrate and is instead a function of the nature of interactions (e.g.~hard, repulsive, attractive) between the liquid atoms and the substrate~\cite{AbrahamJCP1978,DavisJCP1992}. We therefore believe that our key findings will not change if a different type of substrate, or a different facet of the fcc crystal, is used.

As discussed in several publications~\cite{SchofieldProcRSocLondA1982,BausPRL1990,TersoffPRL1991,  BlokhuisJCP1992, TestaJCP2006}, an ambiguity exists in the definition of the position-dependent stress tensor in inhomogeneous systems. In thin films studied in this work, however, all thermodynamic and kinetic properties, including the stress tensor, are functions of $z$ only, and, as shown in Appendix~\ref{appendix:ambiguity}, all components of the undetermined stress tensor are not functions of $z$. Since we are not concerned about the precise values of the stress tensor, and instead, focus on its oscillations across the film, none of our results will be affected by this ambiguity.

\acknowledgments
P.G.D. gratefully acknowledges the support of the National Science
Foundation (Grant No. CHE-1213343). A part of the calculations were performed on the Terascale Infrastructure for Groundbreaking Research in Engineering and Science (TIGRESS) at Princeton University. We gratefully acknowledge R. A. Priestley, J. Dyre, Y. Guo and Z. Shi for useful discussions. 

\appendix
\section{\label{appendix:model}A Simple Thermodynamic Model to Understand why B Atoms Tend to Avoid the Interfacial Regions}

In order to understand why an A-rich interface will be more stable, we construct a simple thermodynamic model that is based on a mean-field approximation. We partition the film into two non-interacting regions that are internally well-mixed and that can freely exchange particles between each other. We also employ a 'pseudo-lattice approach` and only consider the contributions of nearest-neighbor pairs to the internal energy of the system. With these simple assumptions, the free energy of the system can be expressed as:
\begin{eqnarray}
A_{\text{total}} &=& f_iA_i + (1-f_i)A_b \\
A_i &=& \tfrac{\mathfrak{z}}{4}\left[\tfrac32x_{B,i}^2-x_{B,i}-1\right]\notag\\&&+T^*\left[
x_{B,i}\log x_{B,i}+(1-x_{B,i})\log(1-x_{B,i})\right]\notag\\&& \label{eq:A_interfacial} \\
A_b &=& \tfrac{\mathfrak{z}}{2}\left[\tfrac32x_{B,b}^2-x_{B,b}-1\right]\notag\\&&+T^*\left[
x_{B,b}\log x_{B,b}+(1-x_{B,b})\log(1-x_{B,b})\right]\notag\\&&\label{eq:A_bulk}
\end{eqnarray}
where $f_i$ is the fraction of particles that reside in the interfacial region, $x_{B,b}$ and $x_{B,i}$ are the mole fractions of B in the bulk and in the interfacial region, and $\mathfrak{z}$ is the average coordination number. The additional $\tfrac12$ factor in Eq.~(\ref{eq:A_interfacial}) is to account for the nearest-neighbor interactions lost in the interface. By taking $f_i=\frac18$ and $\mathfrak{z}=12$ and by noting that $f_ix_{B,i}+(1-f_i)x_{B,b}=x_{B,\text{total}}=\tfrac15$, we obtain $A_{\text{total}}, A_b$ and $A_i$ vs. $x_{B,i}$ for different temperatures. As can be seen in Fig.~\ref{fig:helmapprox}, the total free energy of this model is minimized if $x_{B,i}<x_{B,\text{total}}$. Increasing the fraction of B particles in the bulk region is both energetically and entropically favorable, however this favorability is offset by an increase in the energy and decrease in the entropy of the interfacial region. The interplay between these two competing effects leads to an interfacial region that has a smaller fraction of B atoms than the bulk, something that had been observed in earlier studies of free-standing thin films as well~\cite{ShiJCP2011}. 

\allowdisplaybreaks[1]

\section{Ambiguity in the Definition of the Stress Tensor
\label{appendix:ambiguity}
}
As mentioned in Ref.~\cite{TestaJCP2006}, the stress tensor given by Eq.~(\ref{eq:stress}) is ambiguous upon the addition of a symmetrized divergenceless traceless tensor $\mathcal{T}$. In thin films considered in this work, $\mathcal{T}$ is a function of $z$ only and $\partial\mathcal{T}/\partial{x}=\partial\mathcal{T}/\partial{y}=\textbf{0}$. Therefore, $\nabla\cdot\mathcal{T}=\textbf{0}$ implies that:
\begin{eqnarray}
	\left[\nabla\cdot\mathcal{T}\right]_x = \frac{\partial\mathcal{T}^{xz}}{\partial z} = 0 \notag\\
	\left[\nabla\cdot\mathcal{T}\right]_y = \frac{\partial\mathcal{T}^{yz}}{\partial z} = 0 \notag\\
	\left[\nabla\cdot\mathcal{T}\right]_z = \frac{\partial\mathcal{T}^{zz}}{\partial z} = 0 \notag
\end{eqnarray}
and $\mathcal{T}$ is thus a constant traceless tensor since $\partial\mathcal{T}/\partial{z}=0$. Consequently, the position dependence of the stress profiles will not be affected by the addition of the constant $\mathcal{T}$ tensor, and all the observed oscillations in lateral and normal stress will be unchanged.

\bibliographystyle{apsrev}

\begin{thebibliography}{0}
\expandafter\ifx\csname natexlab\endcsname\relax\def\natexlab#1{#1}\fi
\expandafter\ifx\csname bibnamefont\endcsname\relax
  \def\bibnamefont#1{#1}\fi
\expandafter\ifx\csname bibfnamefont\endcsname\relax
  \def\bibfnamefont#1{#1}\fi
\expandafter\ifx\csname citenamefont\endcsname\relax
  \def\citenamefont#1{#1}\fi
\expandafter\ifx\csname url\endcsname\relax
  \def\url#1{\texttt{#1}}\fi
\expandafter\ifx\csname urlprefix\endcsname\relax\def\urlprefix{URL }\fi
\providecommand{\bibinfo}[2]{#2}
\providecommand{\eprint}[2][]{\url{#2}}

\end{thebibliography}


\begin{thebibliography}{97}
\expandafter\ifx\csname natexlab\endcsname\relax\def\natexlab#1{#1}\fi
\expandafter\ifx\csname bibnamefont\endcsname\relax
  \def\bibnamefont#1{#1}\fi
\expandafter\ifx\csname bibfnamefont\endcsname\relax
  \def\bibfnamefont#1{#1}\fi
\expandafter\ifx\csname citenamefont\endcsname\relax
  \def\citenamefont#1{#1}\fi
\expandafter\ifx\csname url\endcsname\relax
  \def\url#1{\texttt{#1}}\fi
\expandafter\ifx\csname urlprefix\endcsname\relax\def\urlprefix{URL }\fi
\providecommand{\bibinfo}[2]{#2}
\providecommand{\eprint}[2][]{\url{#2}}

\bibitem[{\citenamefont{Schaller}(1997)}]{IEEESpectrum1997}
\bibinfo{author}{\bibfnamefont{R.~R.} \bibnamefont{Schaller}},
  \bibinfo{journal}{IEEE Spectrum} \textbf{\bibinfo{volume}{34}},
  \bibinfo{pages}{53} (\bibinfo{year}{1997}),
  \urlprefix\url{http://dx.doi.org/10.1109/6.591665}.

\bibitem[{\citenamefont{Volokitin et~al.}(1996)\citenamefont{Volokitin, Sinzig,
  Jongh, Schmid, Vargaftik, and Moiseevi}}]{MoiseeviNature1996}
\bibinfo{author}{\bibfnamefont{Y.}~\bibnamefont{Volokitin}},
  \bibinfo{author}{\bibfnamefont{J.}~\bibnamefont{Sinzig}},
  \bibinfo{author}{\bibfnamefont{L.~J.~D.} \bibnamefont{Jongh}},
  \bibinfo{author}{\bibfnamefont{G.}~\bibnamefont{Schmid}},
  \bibinfo{author}{\bibfnamefont{M.~N.} \bibnamefont{Vargaftik}},
  \bibnamefont{and} \bibinfo{author}{\bibfnamefont{I.~I.}
  \bibnamefont{Moiseevi}}, \bibinfo{journal}{Nature}
  \textbf{\bibinfo{volume}{384}}, \bibinfo{pages}{621} (\bibinfo{year}{1996}),
  \urlprefix\url{http://dx.doi.org/10.1038/384621a0}.

\bibitem[{\citenamefont{Daniel and Astruc}(2004)}]{DanielChemRev2004}
\bibinfo{author}{\bibfnamefont{M.-C.} \bibnamefont{Daniel}} \bibnamefont{and}
  \bibinfo{author}{\bibfnamefont{D.}~\bibnamefont{Astruc}},
  \bibinfo{journal}{Chem. Rev.} \textbf{\bibinfo{volume}{104}},
  \bibinfo{pages}{293} (\bibinfo{year}{2004}),
  \urlprefix\url{http://dx.doi.org/10.1021/cr030698}.

\bibitem[{\citenamefont{Marinica et~al.}(2012)\citenamefont{Marinica, Kazansky,
  Nordlander, Aizpurua, and Borisov}}]{MarinicaNanoLetters2012}
\bibinfo{author}{\bibfnamefont{D.}~\bibnamefont{Marinica}},
  \bibinfo{author}{\bibfnamefont{A.}~\bibnamefont{Kazansky}},
  \bibinfo{author}{\bibfnamefont{P.}~\bibnamefont{Nordlander}},
  \bibinfo{author}{\bibfnamefont{J.}~\bibnamefont{Aizpurua}}, \bibnamefont{and}
  \bibinfo{author}{\bibfnamefont{A.~G.} \bibnamefont{Borisov}},
  \bibinfo{journal}{Nano Lett.} \textbf{\bibinfo{volume}{12}},
  \bibinfo{pages}{1333} (\bibinfo{year}{2012}),
  \urlprefix\url{http://dx.doi.org/10.1021/nl300269c}.

\bibitem[{\citenamefont{Minton}(2001)}]{MintonJBC2001}
\bibinfo{author}{\bibfnamefont{A.~P.} \bibnamefont{Minton}},
  \bibinfo{journal}{J, Biol. Chem.} \textbf{\bibinfo{volume}{276}},
  \bibinfo{pages}{10577} (\bibinfo{year}{2001}),
  \urlprefix\url{http://dx.doi.org/10.1074/jbc.R100005200}.

\bibitem[{\citenamefont{Ando and Skolnick}(2010)}]{SkolnickPNAS2010}
\bibinfo{author}{\bibfnamefont{T.}~\bibnamefont{Ando}} \bibnamefont{and}
  \bibinfo{author}{\bibfnamefont{J.}~\bibnamefont{Skolnick}},
  \bibinfo{journal}{Proc. Natl. Acad. Sci. USA} \textbf{\bibinfo{volume}{107}},
  \bibinfo{pages}{18457} (\bibinfo{year}{2010}),
  \urlprefix\url{http://dx.doi.org/10.1073/pnas.1011354107}.

\bibitem[{\citenamefont{Wang et~al.}(2012)\citenamefont{Wang, Wu, Zhang, Liu,
  Yang, and Feng}}]{WangJHydrol2012}
\bibinfo{author}{\bibfnamefont{J.}~\bibnamefont{Wang}},
  \bibinfo{author}{\bibfnamefont{Y.}~\bibnamefont{Wu}},
  \bibinfo{author}{\bibfnamefont{X.}~\bibnamefont{Zhang}},
  \bibinfo{author}{\bibfnamefont{Y.}~\bibnamefont{Liu}},
  \bibinfo{author}{\bibfnamefont{T.}~\bibnamefont{Yang}}, \bibnamefont{and}
  \bibinfo{author}{\bibfnamefont{B.}~\bibnamefont{Feng}}, \bibinfo{journal}{J.
  Hydrol.} \textbf{\bibinfo{volume}{464--465}}, \bibinfo{pages}{328}
  (\bibinfo{year}{2012}),
  \urlprefix\url{http://dx.doi.org/10.1016/j.jhydrol.2012.07.018}.

\bibitem[{\citenamefont{Clennell et~al.}(1999)\citenamefont{Clennell, Hovland,
  Booth, Henry, and Winters}}]{ChannelJGR1999}
\bibinfo{author}{\bibfnamefont{M.~B.} \bibnamefont{Clennell}},
  \bibinfo{author}{\bibfnamefont{M.}~\bibnamefont{Hovland}},
  \bibinfo{author}{\bibfnamefont{J.~S.} \bibnamefont{Booth}},
  \bibinfo{author}{\bibfnamefont{P.}~\bibnamefont{Henry}}, \bibnamefont{and}
  \bibinfo{author}{\bibfnamefont{W.~J.} \bibnamefont{Winters}},
  \bibinfo{journal}{J. Geophys. Res.} \textbf{\bibinfo{volume}{104}},
  \bibinfo{pages}{22985} (\bibinfo{year}{1999}),
  \urlprefix\url{http://dx.doi.org/10.1029/1999JB900175}.

\bibitem[{\citenamefont{Westbrook and Illingworth}(2013)}]{WestbrookQJRMS2013}
\bibinfo{author}{\bibfnamefont{C.~D.} \bibnamefont{Westbrook}}
  \bibnamefont{and} \bibinfo{author}{\bibfnamefont{A.~J.}
  \bibnamefont{Illingworth}}, \bibinfo{journal}{Q. J. Roy. Meteor. Soc.}
  (\bibinfo{year}{2013}), \urlprefix\url{http://dx.doi.org/10.1002/qj.2096}.

\bibitem[{\citenamefont{Israelachvili and
  McGuiggan}(1990)}]{IsraelachviliJMatRes1990}
\bibinfo{author}{\bibfnamefont{J.~N.} \bibnamefont{Israelachvili}}
  \bibnamefont{and} \bibinfo{author}{\bibfnamefont{P.~M.}
  \bibnamefont{McGuiggan}}, \bibinfo{journal}{J. Mat. Res.}
  \textbf{\bibinfo{volume}{5}}, \bibinfo{pages}{2232} (\bibinfo{year}{1990}),
  \urlprefix\url{http://dx.doi.org/10.1557/JMR.1990.2223}.

\bibitem[{\citenamefont{Heuberger et~al.}(2001)\citenamefont{Heuberger,
  Z\"{a}ch, and Spencer}}]{SpencerScience2001}
\bibinfo{author}{\bibfnamefont{M.}~\bibnamefont{Heuberger}},
  \bibinfo{author}{\bibfnamefont{M.}~\bibnamefont{Z\"{a}ch}}, \bibnamefont{and}
  \bibinfo{author}{\bibfnamefont{N.~D.} \bibnamefont{Spencer}},
  \bibinfo{journal}{Science} \textbf{\bibinfo{volume}{292}},
  \bibinfo{pages}{905} (\bibinfo{year}{2001}),
  \urlprefix\url{http://dx.doi.org/10.1126/science.1058573}.

\bibitem[{\citenamefont{Bell et~al.}(2003)\citenamefont{Bell, Wang, Iedema, and
  Cowin}}]{BellJACS2003}
\bibinfo{author}{\bibfnamefont{R.~C.} \bibnamefont{Bell}},
  \bibinfo{author}{\bibfnamefont{H.}~\bibnamefont{Wang}},
  \bibinfo{author}{\bibfnamefont{M.~J.} \bibnamefont{Iedema}},
  \bibnamefont{and} \bibinfo{author}{\bibfnamefont{J.~P.} \bibnamefont{Cowin}},
  \bibinfo{journal}{J. Am. Chem. Soc.} pp. \bibinfo{pages}{5176--5185}
  (\bibinfo{year}{2003}), \urlprefix\url{http://dx.doi.org/10.1021/ja0291437}.

\bibitem[{\citenamefont{Zhu et~al.}(2011)\citenamefont{Zhu, Brian, Swallen,
  Straus, Ediger, and Yu}}]{ZhuEdigerPRL2011}
\bibinfo{author}{\bibfnamefont{L.}~\bibnamefont{Zhu}},
  \bibinfo{author}{\bibfnamefont{C.~W.} \bibnamefont{Brian}},
  \bibinfo{author}{\bibfnamefont{S.~F.} \bibnamefont{Swallen}},
  \bibinfo{author}{\bibfnamefont{P.~T.} \bibnamefont{Straus}},
  \bibinfo{author}{\bibfnamefont{M.~D.} \bibnamefont{Ediger}},
  \bibnamefont{and} \bibinfo{author}{\bibfnamefont{L.}~\bibnamefont{Yu}},
  \bibinfo{journal}{Phys. Rev. Lett.} \textbf{\bibinfo{volume}{106}},
  \bibinfo{pages}{256103} (\bibinfo{year}{2011}),
  \urlprefix\url{http://dx.doi.org/10.1103/PhysRevLett.106.256103}.

\bibitem[{\citenamefont{Khan et~al.}(2010)\citenamefont{Khan, Matei, Patil, and
  Hoffmann}}]{HoffmanPRL2010}
\bibinfo{author}{\bibfnamefont{S.~H.} \bibnamefont{Khan}},
  \bibinfo{author}{\bibfnamefont{G.}~\bibnamefont{Matei}},
  \bibinfo{author}{\bibfnamefont{S.}~\bibnamefont{Patil}}, \bibnamefont{and}
  \bibinfo{author}{\bibfnamefont{P.~M.} \bibnamefont{Hoffmann}},
  \bibinfo{journal}{Phys. Rev. Lett.} \textbf{\bibinfo{volume}{105}},
  \bibinfo{pages}{106101} (\bibinfo{year}{2010}),
  \urlprefix\url{http://dx.doi.org/10.1103/PhysRevLett.105.106101}.

\bibitem[{\citenamefont{Chai et~al.}(2014)\citenamefont{Chai, Salez, McGraw,
  Benzaquen, Dalnoki-Veress, Rapha\"{e}l, and Forrest}}]{ForestScience2014}
\bibinfo{author}{\bibfnamefont{Y.}~\bibnamefont{Chai}},
  \bibinfo{author}{\bibfnamefont{T.}~\bibnamefont{Salez}},
  \bibinfo{author}{\bibfnamefont{J.~D.} \bibnamefont{McGraw}},
  \bibinfo{author}{\bibfnamefont{M.}~\bibnamefont{Benzaquen}},
  \bibinfo{author}{\bibfnamefont{K.}~\bibnamefont{Dalnoki-Veress}},
  \bibinfo{author}{\bibfnamefont{E.}~\bibnamefont{Rapha\"{e}l}},
  \bibnamefont{and} \bibinfo{author}{\bibfnamefont{J.~A.}
  \bibnamefont{Forrest}}, \bibinfo{journal}{Science}
  \textbf{\bibinfo{volume}{343}}, \bibinfo{pages}{994} (\bibinfo{year}{2014}),
  \urlprefix\url{http://dx.doi.org/10.1126/science.1244845}.

\bibitem[{\citenamefont{Bitsanis and Hadziioannou}(1990)}]{HadziioannouJCP1990}
\bibinfo{author}{\bibfnamefont{D.~P.~I.} \bibnamefont{Bitsanis}}
  \bibnamefont{and}
  \bibinfo{author}{\bibfnamefont{G.}~\bibnamefont{Hadziioannou}},
  \bibinfo{journal}{J. Chem. Phys.} \textbf{\bibinfo{volume}{92}},
  \bibinfo{pages}{3827} (\bibinfo{year}{1990}),
  \urlprefix\url{http://dx.doi.org/10.1063/1.457840}.

\bibitem[{\citenamefont{Somers and Davis}(1992)}]{DavisJCP1992}
\bibinfo{author}{\bibfnamefont{S.~A.} \bibnamefont{Somers}} \bibnamefont{and}
  \bibinfo{author}{\bibfnamefont{H.~T.} \bibnamefont{Davis}},
  \bibinfo{journal}{J. Chem. Phys.} \textbf{\bibinfo{volume}{96}},
  \bibinfo{pages}{5389} (\bibinfo{year}{1992}),
  \urlprefix\url{http://dx.doi.org/10.1063/1.462724}.

\bibitem[{\citenamefont{Lee and Rossky}(1994)}]{RosskyJCP1994}
\bibinfo{author}{\bibfnamefont{S.~H.} \bibnamefont{Lee}} \bibnamefont{and}
  \bibinfo{author}{\bibfnamefont{P.~J.} \bibnamefont{Rossky}},
  \bibinfo{journal}{J. Chem. Phys.} \textbf{\bibinfo{volume}{100}},
  \bibinfo{pages}{3334} (\bibinfo{year}{1994}),
  \urlprefix\url{http://dx.doi.org/10.1063/1.466425}.

\bibitem[{\citenamefont{Marrink and Berendsen}(1994)}]{BerendsenJPhysChem1994}
\bibinfo{author}{\bibfnamefont{S.-J.} \bibnamefont{Marrink}} \bibnamefont{and}
  \bibinfo{author}{\bibfnamefont{H.~J.~C.} \bibnamefont{Berendsen}},
  \bibinfo{journal}{J. Phys. Chem.} \textbf{\bibinfo{volume}{98}},
  \bibinfo{pages}{4155} (\bibinfo{year}{1994}),
  \urlprefix\url{http://dx.doi.org/10.1021/j100066a040}.

\bibitem[{\citenamefont{Manias et~al.}(1996)\citenamefont{Manias, Hadziioannou,
  and ten Brinke}}]{BrinkeLangmuir1995}
\bibinfo{author}{\bibfnamefont{E.}~\bibnamefont{Manias}},
  \bibinfo{author}{\bibfnamefont{G.}~\bibnamefont{Hadziioannou}},
  \bibnamefont{and} \bibinfo{author}{\bibfnamefont{G.}~\bibnamefont{ten
  Brinke}}, \bibinfo{journal}{Langmuir} \textbf{\bibinfo{volume}{12}},
  \bibinfo{pages}{4587} (\bibinfo{year}{1996}),
  \urlprefix\url{http://dx.doi.org/10.1021/la950902r}.

\bibitem[{\citenamefont{Gao et~al.}(1997)\citenamefont{Gao, Luedtke, and
  Landman}}]{LandmanPRL1997}
\bibinfo{author}{\bibfnamefont{J.}~\bibnamefont{Gao}},
  \bibinfo{author}{\bibfnamefont{W.~D.} \bibnamefont{Luedtke}},
  \bibnamefont{and} \bibinfo{author}{\bibfnamefont{U.}~\bibnamefont{Landman}},
  \bibinfo{journal}{Phys. Rev. Lett.} \textbf{\bibinfo{volume}{79}},
  \bibinfo{pages}{705} (\bibinfo{year}{1997}),
  \urlprefix\url{http://dx.doi.org/10.1103/PhysRevLett.79.705}.

\bibitem[{\citenamefont{Porcheron et~al.}(2002)\citenamefont{Porcheron, Schoen,
  and Fuchs}}]{FuchsJCP2002}
\bibinfo{author}{\bibfnamefont{F.}~\bibnamefont{Porcheron}},
  \bibinfo{author}{\bibfnamefont{M.}~\bibnamefont{Schoen}}, \bibnamefont{and}
  \bibinfo{author}{\bibfnamefont{A.~H.} \bibnamefont{Fuchs}},
  \bibinfo{journal}{J. Chem. Phys.} \textbf{\bibinfo{volume}{116}},
  \bibinfo{pages}{5816} (\bibinfo{year}{2002}),
  \urlprefix\url{http://dx.doi.org/10.1063/1.1453968}.

\bibitem[{\citenamefont{Teboul and Simionesco}(2002)}]{TeboulJPhysCondMat2002}
\bibinfo{author}{\bibfnamefont{V.}~\bibnamefont{Teboul}} \bibnamefont{and}
  \bibinfo{author}{\bibfnamefont{C.~A.} \bibnamefont{Simionesco}},
  \bibinfo{journal}{J. Phys.: Condens. Matter} \textbf{\bibinfo{volume}{14}},
  \bibinfo{pages}{5699} (\bibinfo{year}{2002}),
  \urlprefix\url{http://dx.doi.org/10.1088/0953-8984/14/23/304}.

\bibitem[{\citenamefont{Lan\c{c}on et~al.}(2002)\citenamefont{Lan\c{c}on,
  Batrouni, Lobry, and Ostrowsky}}]{OstrowskyPhysicaA2002}
\bibinfo{author}{\bibfnamefont{P.}~\bibnamefont{Lan\c{c}on}},
  \bibinfo{author}{\bibfnamefont{G.}~\bibnamefont{Batrouni}},
  \bibinfo{author}{\bibfnamefont{L.}~\bibnamefont{Lobry}}, \bibnamefont{and}
  \bibinfo{author}{\bibfnamefont{N.}~\bibnamefont{Ostrowsky}},
  \bibinfo{journal}{Physica A} \textbf{\bibinfo{volume}{304}},
  \bibinfo{pages}{65} (\bibinfo{year}{2002}),
  \urlprefix\url{http://dx.doi.org/10.1016/S0378-4371(01)00510-6}.

\bibitem[{\citenamefont{Liu et~al.}(2004)\citenamefont{Liu, Harder, and
  Berne}}]{BerneJPhysChemB2004}
\bibinfo{author}{\bibfnamefont{P.}~\bibnamefont{Liu}},
  \bibinfo{author}{\bibfnamefont{E.}~\bibnamefont{Harder}}, \bibnamefont{and}
  \bibinfo{author}{\bibfnamefont{B.~J.} \bibnamefont{Berne}},
  \bibinfo{journal}{J. Phys. Chem. B} \textbf{\bibinfo{volume}{108}},
  \bibinfo{pages}{6595} (\bibinfo{year}{2004}),
  \urlprefix\url{http://dx.doi.org/10.1021/jp0375057}.

\bibitem[{\citenamefont{Desai et~al.}(2005)\citenamefont{Desai, Keblinski, and
  Kumar}}]{KumarJCP2005}
\bibinfo{author}{\bibfnamefont{T.}~\bibnamefont{Desai}},
  \bibinfo{author}{\bibfnamefont{P.}~\bibnamefont{Keblinski}},
  \bibnamefont{and} \bibinfo{author}{\bibfnamefont{S.~K.} \bibnamefont{Kumar}},
  \bibinfo{journal}{J. Chem. Phys.} \textbf{\bibinfo{volume}{122}},
  \bibinfo{pages}{134910} (\bibinfo{year}{2005}),
  \urlprefix\url{http://dx.doi.org/10.1063/1.1874852}.

\bibitem[{\citenamefont{Shi et~al.}(2011)\citenamefont{Shi, Debenedetti, and
  Stillinger}}]{ShiJCP2011}
\bibinfo{author}{\bibfnamefont{Z.}~\bibnamefont{Shi}},
  \bibinfo{author}{\bibfnamefont{P.~G.} \bibnamefont{Debenedetti}},
  \bibnamefont{and} \bibinfo{author}{\bibfnamefont{F.~H.}
  \bibnamefont{Stillinger}}, \bibinfo{journal}{J. Chem. Phys.}
  \textbf{\bibinfo{volume}{134}}, \bibinfo{pages}{114524}
  (\bibinfo{year}{2011}), \urlprefix\url{http://dx.doi.org/10.1063/1.3565480}.

\bibitem[{\citenamefont{de~Beer et~al.}(2012)\citenamefont{de~Beer, den Otter,
  van~den Ende, Briels, and Mugele}}]{deBeerEPL2012}
\bibinfo{author}{\bibfnamefont{S.}~\bibnamefont{de~Beer}},
  \bibinfo{author}{\bibfnamefont{W.~K.} \bibnamefont{den Otter}},
  \bibinfo{author}{\bibfnamefont{D.}~\bibnamefont{van~den Ende}},
  \bibinfo{author}{\bibfnamefont{W.~J.} \bibnamefont{Briels}},
  \bibnamefont{and} \bibinfo{author}{\bibfnamefont{F.}~\bibnamefont{Mugele}},
  \bibinfo{journal}{Europhys. Lett.} \textbf{\bibinfo{volume}{97}},
  \bibinfo{pages}{46001} (\bibinfo{year}{2012}),
  \urlprefix\url{http://dx.doi.org/10.1209/0295-5075/97/46001}.

\bibitem[{\citenamefont{Phan et~al.}(2012)\citenamefont{Phan, Ho, Cole, and
  Striolo}}]{StrioloJPCC2012}
\bibinfo{author}{\bibfnamefont{A.}~\bibnamefont{Phan}},
  \bibinfo{author}{\bibfnamefont{T.~A.} \bibnamefont{Ho}},
  \bibinfo{author}{\bibfnamefont{D.~R.} \bibnamefont{Cole}}, \bibnamefont{and}
  \bibinfo{author}{\bibfnamefont{A.}~\bibnamefont{Striolo}},
  \bibinfo{journal}{J. Phys. Chem. C} \textbf{\bibinfo{volume}{116}},
  \bibinfo{pages}{15962} (\bibinfo{year}{2012}),
  \urlprefix\url{http://dx.doi.org/10.1021/jp300679v}.

\bibitem[{\citenamefont{Christenson}(2001)}]{ChristensonJPCM2001}
\bibinfo{author}{\bibfnamefont{H.~K.} \bibnamefont{Christenson}},
  \bibinfo{journal}{J. Phys.-Condens. Mat.} \textbf{\bibinfo{volume}{13}},
  \bibinfo{pages}{R95} (\bibinfo{year}{2001}),
  \urlprefix\url{http://dx.doi.org/10.1088/0953-8984/13/11/201}.

\bibitem[{\citenamefont{Taschin et~al.}(2010)\citenamefont{Taschin, Cucini,
  Bartolini, and Torre}}]{TaschinEPL2010}
\bibinfo{author}{\bibfnamefont{A.}~\bibnamefont{Taschin}},
  \bibinfo{author}{\bibfnamefont{R.}~\bibnamefont{Cucini}},
  \bibinfo{author}{\bibfnamefont{P.}~\bibnamefont{Bartolini}},
  \bibnamefont{and} \bibinfo{author}{\bibfnamefont{R.}~\bibnamefont{Torre}},
  \bibinfo{journal}{Europhys. Lett.} \textbf{\bibinfo{volume}{92}},
  \bibinfo{pages}{26005} (\bibinfo{year}{2010}),
  \urlprefix\url{http://dx.doi.org/10.1209/0295-5075/92/26005}.

\bibitem[{\citenamefont{Richert}(2011)}]{RichertARPC2010}
\bibinfo{author}{\bibfnamefont{R.}~\bibnamefont{Richert}},
  \bibinfo{journal}{Ann. Rev. Phys. Chem.} \textbf{\bibinfo{volume}{62}},
  \bibinfo{pages}{65} (\bibinfo{year}{2011}),
  \urlprefix\url{http://dx.doi.org/10.1146/annurev-physchem-032210-103343}.

\bibitem[{\citenamefont{Zhang et~al.}(2011)\citenamefont{Zhang, Guo, and
  Priestley}}]{PriestleyMacromolecules2011}
\bibinfo{author}{\bibfnamefont{C.}~\bibnamefont{Zhang}},
  \bibinfo{author}{\bibfnamefont{Y.}~\bibnamefont{Guo}}, \bibnamefont{and}
  \bibinfo{author}{\bibfnamefont{R.~D.} \bibnamefont{Priestley}},
  \bibinfo{journal}{Macromolecules} \textbf{\bibinfo{volume}{44}},
  \bibinfo{pages}{4001} (\bibinfo{year}{2011}),
  \urlprefix\url{http://dx.doi.org/10.1021/ma1026862}.

\bibitem[{\citenamefont{Oguni et~al.}(2011)\citenamefont{Oguni, Kanke, Nagoe,
  and Namba}}]{OguniJPCB2011}
\bibinfo{author}{\bibfnamefont{M.}~\bibnamefont{Oguni}},
  \bibinfo{author}{\bibfnamefont{Y.}~\bibnamefont{Kanke}},
  \bibinfo{author}{\bibfnamefont{A.}~\bibnamefont{Nagoe}}, \bibnamefont{and}
  \bibinfo{author}{\bibfnamefont{S.}~\bibnamefont{Namba}}, \bibinfo{journal}{J.
  Phys. Chem. B} \textbf{\bibinfo{volume}{115}}, \bibinfo{pages}{14023}
  (\bibinfo{year}{2011}), \urlprefix\url{http://dx.doi.org/10.1021/jp2034032}.

\bibitem[{\citenamefont{Johnston et~al.}(2010)\citenamefont{Johnston,
  Kastelowitz, and Molinero}}]{MolineroJCP2010}
\bibinfo{author}{\bibfnamefont{J.~C.} \bibnamefont{Johnston}},
  \bibinfo{author}{\bibfnamefont{N.}~\bibnamefont{Kastelowitz}},
  \bibnamefont{and} \bibinfo{author}{\bibfnamefont{V.}~\bibnamefont{Molinero}},
  \bibinfo{journal}{J. Chem. Phys.} \textbf{\bibinfo{volume}{133}},
  \bibinfo{pages}{154516} (\bibinfo{year}{2010}),
  \urlprefix\url{http://dx.doi.org/10.1063/1.3499323}.

\bibitem[{\citenamefont{Fayer and Levinger}(2010)}]{LevingerARAC2010}
\bibinfo{author}{\bibfnamefont{M.~D.} \bibnamefont{Fayer}} \bibnamefont{and}
  \bibinfo{author}{\bibfnamefont{N.~E.} \bibnamefont{Levinger}},
  \bibinfo{journal}{Ann. Rev. Anal. Chem.} \textbf{\bibinfo{volume}{3}},
  \bibinfo{pages}{89} (\bibinfo{year}{2010}),
  \urlprefix\url{http://dx.doi.org/10.1146/annurev-anchem-070109-103410}.

\bibitem[{\citenamefont{J.-P.~Hirvonen
  et~al.}(1997)\citenamefont{J.-P.~Hirvonen, Kaukonen, Nieminen, and
  Scheibe}}]{HirvonenJAppPhys1997}
\bibinfo{author}{\bibfnamefont{J.~K.} \bibnamefont{J.-P.~Hirvonen}},
  \bibinfo{author}{\bibfnamefont{M.}~\bibnamefont{Kaukonen}},
  \bibinfo{author}{\bibfnamefont{R.}~\bibnamefont{Nieminen}}, \bibnamefont{and}
  \bibinfo{author}{\bibfnamefont{H.-J.} \bibnamefont{Scheibe}},
  \bibinfo{journal}{J. Appl. Phys.} \textbf{\bibinfo{volume}{81}},
  \bibinfo{pages}{7248} (\bibinfo{year}{1997}),
  \urlprefix\url{http://dx.doi.org/10.1063/1.365322}.

\bibitem[{\citenamefont{Chan et~al.}(2009)\citenamefont{Chan, Page, Im, Patton,
  Huang, and Stafford}}]{StaffordSoftMatter2009}
\bibinfo{author}{\bibfnamefont{E.~P.} \bibnamefont{Chan}},
  \bibinfo{author}{\bibfnamefont{K.~A.} \bibnamefont{Page}},
  \bibinfo{author}{\bibfnamefont{S.~H.} \bibnamefont{Im}},
  \bibinfo{author}{\bibfnamefont{D.~L.} \bibnamefont{Patton}},
  \bibinfo{author}{\bibfnamefont{R.}~\bibnamefont{Huang}}, \bibnamefont{and}
  \bibinfo{author}{\bibfnamefont{C.~M.} \bibnamefont{Stafford}},
  \bibinfo{journal}{Soft Matter} \textbf{\bibinfo{volume}{5}},
  \bibinfo{pages}{4638} (\bibinfo{year}{2009}),
  \urlprefix\url{http://dx.doi.org/10.1039/B916207K}.

\bibitem[{\citenamefont{Rafiq et~al.}(2010)\citenamefont{Rafiq, Yadav, and
  Sen}}]{RafiqJPCB2010}
\bibinfo{author}{\bibfnamefont{S.}~\bibnamefont{Rafiq}},
  \bibinfo{author}{\bibfnamefont{R.}~\bibnamefont{Yadav}}, \bibnamefont{and}
  \bibinfo{author}{\bibfnamefont{P.}~\bibnamefont{Sen}}, \bibinfo{journal}{J.
  Phys. Chem. B} \textbf{\bibinfo{volume}{114}}, \bibinfo{pages}{13988}
  (\bibinfo{year}{2010}), \urlprefix\url{http://dx.doi.org/10.1021/jp1037238}.

\bibitem[{\citenamefont{Bocquet and Barrat}(1995)}]{BarratEPL1995}
\bibinfo{author}{\bibfnamefont{L.}~\bibnamefont{Bocquet}} \bibnamefont{and}
  \bibinfo{author}{\bibfnamefont{J.-L.} \bibnamefont{Barrat}},
  \bibinfo{journal}{Europhys. Lett.} \textbf{\bibinfo{volume}{31}},
  \bibinfo{pages}{455} (\bibinfo{year}{1995}),
  \urlprefix\url{http://dx.doi.org/10.1209/0295-5075/31/8/006}.

\bibitem[{\citenamefont{He et~al.}(2013)\citenamefont{He, Khorasani, Retterer,
  Thomas, Conrad, and Krishnamoorti}}]{KrishnamoortiACSNano2013}
\bibinfo{author}{\bibfnamefont{K.}~\bibnamefont{He}},
  \bibinfo{author}{\bibfnamefont{F.~B.} \bibnamefont{Khorasani}},
  \bibinfo{author}{\bibfnamefont{S.~T.} \bibnamefont{Retterer}},
  \bibinfo{author}{\bibfnamefont{D.~K.} \bibnamefont{Thomas}},
  \bibinfo{author}{\bibfnamefont{J.~C.} \bibnamefont{Conrad}},
  \bibnamefont{and}
  \bibinfo{author}{\bibfnamefont{R.}~\bibnamefont{Krishnamoorti}},
  \bibinfo{journal}{ACS Nano} \textbf{\bibinfo{volume}{7}},
  \bibinfo{pages}{5122} (\bibinfo{year}{2013}),
  \urlprefix\url{http://dx.doi.org/10.1021/nn4007303}.

\bibitem[{\citenamefont{Swallen et~al.}(2007)\citenamefont{Swallen, Kearns,
  Mapes, Kim, McMahon, Ediger, Wu, Yu, and Satija}}]{EdigerScience2007}
\bibinfo{author}{\bibfnamefont{S.~F.} \bibnamefont{Swallen}},
  \bibinfo{author}{\bibfnamefont{K.~L.} \bibnamefont{Kearns}},
  \bibinfo{author}{\bibfnamefont{M.~K.} \bibnamefont{Mapes}},
  \bibinfo{author}{\bibfnamefont{Y.~S.} \bibnamefont{Kim}},
  \bibinfo{author}{\bibfnamefont{R.~J.} \bibnamefont{McMahon}},
  \bibinfo{author}{\bibfnamefont{M.~D.} \bibnamefont{Ediger}},
  \bibinfo{author}{\bibfnamefont{T.}~\bibnamefont{Wu}},
  \bibinfo{author}{\bibfnamefont{L.}~\bibnamefont{Yu}}, \bibnamefont{and}
  \bibinfo{author}{\bibfnamefont{S.}~\bibnamefont{Satija}},
  \bibinfo{journal}{Science} \textbf{\bibinfo{volume}{315}},
  \bibinfo{pages}{353} (\bibinfo{year}{2007}),
  \urlprefix\url{http://dx.doi.org/10.1126/science.1135795}.

\bibitem[{\citenamefont{Moynihan et~al.}(1974)\citenamefont{Moynihan, Easteal,
  Wilder, and Tucker}}]{TuckerJCP1974}
\bibinfo{author}{\bibfnamefont{C.~T.} \bibnamefont{Moynihan}},
  \bibinfo{author}{\bibfnamefont{A.~J.} \bibnamefont{Easteal}},
  \bibinfo{author}{\bibfnamefont{J.}~\bibnamefont{Wilder}}, \bibnamefont{and}
  \bibinfo{author}{\bibfnamefont{J.}~\bibnamefont{Tucker}},
  \bibinfo{journal}{J. Chem. Phys.} \textbf{\bibinfo{volume}{78}},
  \bibinfo{pages}{2673} (\bibinfo{year}{1974}),
  \urlprefix\url{http://dx.doi.org/10.1021/j100619a008}.

\bibitem[{\citenamefont{Debenedetti and
  Stillinger}(2001)}]{DebenedettiNature2001}
\bibinfo{author}{\bibfnamefont{P.~G.} \bibnamefont{Debenedetti}}
  \bibnamefont{and} \bibinfo{author}{\bibfnamefont{F.~H.}
  \bibnamefont{Stillinger}}, \bibinfo{journal}{Nature}
  \textbf{\bibinfo{volume}{410}}, \bibinfo{pages}{259} (\bibinfo{year}{2001}),
  \urlprefix\url{http://dx.doi.org/10.1038/35065704}.

\bibitem[{\citenamefont{Singh et~al.}(2013)\citenamefont{Singh, Ediger, and
  de~Pablo}}]{SinghNatMater2013}
\bibinfo{author}{\bibfnamefont{S.}~\bibnamefont{Singh}},
  \bibinfo{author}{\bibfnamefont{M.~D.} \bibnamefont{Ediger}},
  \bibnamefont{and} \bibinfo{author}{\bibfnamefont{J.~J.}
  \bibnamefont{de~Pablo}}, \bibinfo{journal}{Nat. Mater.}
  \textbf{\bibinfo{volume}{12}}, \bibinfo{pages}{139} (\bibinfo{year}{2013}),
  \urlprefix\url{http://dx.doi.org/10.1038/nmat3521}.

\bibitem[{\citenamefont{Swallen et~al.}(2009)\citenamefont{Swallen, Traynor,
  McMahon, Ediger, and Mates}}]{SwallenPRL2009}
\bibinfo{author}{\bibfnamefont{S.~F.} \bibnamefont{Swallen}},
  \bibinfo{author}{\bibfnamefont{K.}~\bibnamefont{Traynor}},
  \bibinfo{author}{\bibfnamefont{R.~J.} \bibnamefont{McMahon}},
  \bibinfo{author}{\bibfnamefont{M.~D.} \bibnamefont{Ediger}},
  \bibnamefont{and} \bibinfo{author}{\bibfnamefont{T.~E.} \bibnamefont{Mates}},
  \bibinfo{journal}{Phys. Rev. Lett.} \textbf{\bibinfo{volume}{102}},
  \bibinfo{pages}{065503} (\bibinfo{year}{2009}),
  \urlprefix\url{http://dx.doi.org/10.1103/PhysRevLett.102.065503}.

\bibitem[{\citenamefont{Ishii et~al.}(2008)\citenamefont{Ishii, Nakayama,
  Hirabayashi, and Moriyama}}]{IshiiCPL2008}
\bibinfo{author}{\bibfnamefont{K.}~\bibnamefont{Ishii}},
  \bibinfo{author}{\bibfnamefont{H.}~\bibnamefont{Nakayama}},
  \bibinfo{author}{\bibfnamefont{S.}~\bibnamefont{Hirabayashi}},
  \bibnamefont{and} \bibinfo{author}{\bibfnamefont{R.}~\bibnamefont{Moriyama}},
  \bibinfo{journal}{Chem. Phys. Lett.} \textbf{\bibinfo{volume}{459}},
  \bibinfo{pages}{109} (\bibinfo{year}{2008}),
  \urlprefix\url{http://dx.doi.org/10.1016/j.cplett.2008.05.050}.

\bibitem[{\citenamefont{Leon-Gutierrez
  et~al.}(2010{\natexlab{a}})\citenamefont{Leon-Gutierrez, Sep\'{u}lveda,
  Garcia, Clavaguera-Mora, and Rodr\'{i}guez-Viejo}}]{GutierrezPCCP2010}
\bibinfo{author}{\bibfnamefont{E.}~\bibnamefont{Leon-Gutierrez}},
  \bibinfo{author}{\bibfnamefont{A.}~\bibnamefont{Sep\'{u}lveda}},
  \bibinfo{author}{\bibfnamefont{G.}~\bibnamefont{Garcia}},
  \bibinfo{author}{\bibfnamefont{M.~T.} \bibnamefont{Clavaguera-Mora}},
  \bibnamefont{and}
  \bibinfo{author}{\bibfnamefont{J.}~\bibnamefont{Rodr\'{i}guez-Viejo}},
  \bibinfo{journal}{Phys. Chem. Chem. Phys.} \textbf{\bibinfo{volume}{12}},
  \bibinfo{pages}{14693} (\bibinfo{year}{2010}{\natexlab{a}}),
  \urlprefix\url{http://dx.doi.org/10.1039/C0CP00208A}.

\bibitem[{\citenamefont{Le\'{o}n-Gutierrez
  et~al.}(2009)\citenamefont{Le\'{o}n-Gutierrez, Garcia, Clavaguera-Mora, and
  Rodr\'{i}guez-Viejo}}]{GutierrezTA2009}
\bibinfo{author}{\bibfnamefont{E.}~\bibnamefont{Le\'{o}n-Gutierrez}},
  \bibinfo{author}{\bibfnamefont{G.}~\bibnamefont{Garcia}},
  \bibinfo{author}{\bibfnamefont{M.}~\bibnamefont{Clavaguera-Mora}},
  \bibnamefont{and}
  \bibinfo{author}{\bibfnamefont{J.}~\bibnamefont{Rodr\'{i}guez-Viejo}},
  \bibinfo{journal}{Thermochim. Acta} \textbf{\bibinfo{volume}{492}},
  \bibinfo{pages}{51} (\bibinfo{year}{2009}),
  \urlprefix\url{http://dx.doi.org/10.1016/j.tca.2009.05.016}.

\bibitem[{\citenamefont{Leon-Gutierrez
  et~al.}(2010{\natexlab{b}})\citenamefont{Leon-Gutierrez, Garcia, Lopeandia,
  Clavaguera-Mora, and Rodr\'{i}guez-Viejo}}]{GutierrezJPCL2010}
\bibinfo{author}{\bibfnamefont{E.}~\bibnamefont{Leon-Gutierrez}},
  \bibinfo{author}{\bibfnamefont{G.}~\bibnamefont{Garcia}},
  \bibinfo{author}{\bibfnamefont{A.~F.} \bibnamefont{Lopeandia}},
  \bibinfo{author}{\bibfnamefont{M.~T.} \bibnamefont{Clavaguera-Mora}},
  \bibnamefont{and}
  \bibinfo{author}{\bibfnamefont{J.}~\bibnamefont{Rodr\'{i}guez-Viejo}},
  \bibinfo{journal}{J. Phys. Chem. Lett.} \textbf{\bibinfo{volume}{1}},
  \bibinfo{pages}{341} (\bibinfo{year}{2010}{\natexlab{b}}),
  \urlprefix\url{http://dx.doi.org/10.1021/jz900178u}.

\bibitem[{\citenamefont{Sep\'{u}lveda et~al.}(2011)\citenamefont{Sep\'{u}lveda,
  Leon-Gutierrez, Gonzalez-Silveira, Rodr\'{i}guez-Tinoco, Clavaguera-Mora, and
  Rodr\'{i}guez-Viejo}}]{SepulvedaPRL2011}
\bibinfo{author}{\bibfnamefont{A.}~\bibnamefont{Sep\'{u}lveda}},
  \bibinfo{author}{\bibfnamefont{E.}~\bibnamefont{Leon-Gutierrez}},
  \bibinfo{author}{\bibfnamefont{M.}~\bibnamefont{Gonzalez-Silveira}},
  \bibinfo{author}{\bibfnamefont{C.}~\bibnamefont{Rodr\'{i}guez-Tinoco}},
  \bibinfo{author}{\bibfnamefont{M.~T.} \bibnamefont{Clavaguera-Mora}},
  \bibnamefont{and}
  \bibinfo{author}{\bibfnamefont{J.}~\bibnamefont{Rodr\'{i}guez-Viejo}},
  \bibinfo{journal}{Phys. Rev. Lett.} \textbf{\bibinfo{volume}{107}},
  \bibinfo{pages}{025901} (\bibinfo{year}{2011}),
  \urlprefix\url{http://dx.doi.org/10.1103/PhysRevLett.107.025901}.

\bibitem[{\citenamefont{Ishii et~al.}(2012)\citenamefont{Ishii, Nakayama, and
  Moriyama}}]{IshiiJPCB2012}
\bibinfo{author}{\bibfnamefont{K.}~\bibnamefont{Ishii}},
  \bibinfo{author}{\bibfnamefont{H.}~\bibnamefont{Nakayama}}, \bibnamefont{and}
  \bibinfo{author}{\bibfnamefont{R.}~\bibnamefont{Moriyama}},
  \bibinfo{journal}{J. Phys. Chem. B} \textbf{\bibinfo{volume}{116}},
  \bibinfo{pages}{935} (\bibinfo{year}{2012}),
  \urlprefix\url{http://dx.doi.org/10.1021/jp209738h}.

\bibitem[{\citenamefont{Whitaker et~al.}(2013)\citenamefont{Whitaker, Scifo,
  and Ediger}}]{WhitakerJPCB2013}
\bibinfo{author}{\bibfnamefont{K.~R.} \bibnamefont{Whitaker}},
  \bibinfo{author}{\bibfnamefont{D.~J.} \bibnamefont{Scifo}}, \bibnamefont{and}
  \bibinfo{author}{\bibfnamefont{M.~D.} \bibnamefont{Ediger}},
  \bibinfo{journal}{J. Phys. Chem. B}  (\bibinfo{year}{2013}),
  \urlprefix\url{http://dx.doi.org/10.1021/jp400960g}.

\bibitem[{\citenamefont{Souda}(2010)}]{SoudaJPCB2010}
\bibinfo{author}{\bibfnamefont{R.}~\bibnamefont{Souda}}, \bibinfo{journal}{J.
  Phys. Chem. B} \textbf{\bibinfo{volume}{114}}, \bibinfo{pages}{11127}
  (\bibinfo{year}{2010}), \urlprefix\url{http://dx.doi.org/10.1021/jp104523h}.

\bibitem[{\citenamefont{Guo et~al.}(2012)\citenamefont{Guo, Morozov, Schneider,
  Chung, Zhang, Waldmann, Yao, Fytas, Arnold, and
  Priestley}}]{PriestleyNatMat2012}
\bibinfo{author}{\bibfnamefont{Y.}~\bibnamefont{Guo}},
  \bibinfo{author}{\bibfnamefont{A.}~\bibnamefont{Morozov}},
  \bibinfo{author}{\bibfnamefont{D.}~\bibnamefont{Schneider}},
  \bibinfo{author}{\bibfnamefont{J.~W.} \bibnamefont{Chung}},
  \bibinfo{author}{\bibfnamefont{C.}~\bibnamefont{Zhang}},
  \bibinfo{author}{\bibfnamefont{M.}~\bibnamefont{Waldmann}},
  \bibinfo{author}{\bibfnamefont{N.}~\bibnamefont{Yao}},
  \bibinfo{author}{\bibfnamefont{G.}~\bibnamefont{Fytas}},
  \bibinfo{author}{\bibfnamefont{C.~B.} \bibnamefont{Arnold}},
  \bibnamefont{and} \bibinfo{author}{\bibfnamefont{R.~D.}
  \bibnamefont{Priestley}}, \bibinfo{journal}{Nat. Mater.}
  \textbf{\bibinfo{volume}{11}}, \bibinfo{pages}{337} (\bibinfo{year}{2012}),
  \urlprefix\url{http://dx.doi.org/10.1038/nmat3234}.

\bibitem[{\citenamefont{Yu et~al.}(2013)\citenamefont{Yu, Luo, and
  Samwer}}]{SamwerAdvMat2013}
\bibinfo{author}{\bibfnamefont{H.-B.} \bibnamefont{Yu}},
  \bibinfo{author}{\bibfnamefont{Y.}~\bibnamefont{Luo}}, \bibnamefont{and}
  \bibinfo{author}{\bibfnamefont{K.}~\bibnamefont{Samwer}},
  \bibinfo{journal}{Adv. Mater.} \textbf{\bibinfo{volume}{25}},
  \bibinfo{pages}{5904} (\bibinfo{year}{2013}),
  \urlprefix\url{http://dx.doi.org/10.1002/adma.201302700}.

\bibitem[{\citenamefont{Zhu and Yu}(2010)}]{ZhuCPL2010}
\bibinfo{author}{\bibfnamefont{L.}~\bibnamefont{Zhu}} \bibnamefont{and}
  \bibinfo{author}{\bibfnamefont{L.}~\bibnamefont{Yu}}, \bibinfo{journal}{Chem.
  Phys. Lett.} \textbf{\bibinfo{volume}{499}}, \bibinfo{pages}{62}
  (\bibinfo{year}{2010}),
  \urlprefix\url{http://dx.doi.org/10.1016/j.cplett.2010.09.010}.

\bibitem[{\citenamefont{Sukas et~al.}(2013)\citenamefont{Sukas, Tiggelaar,
  Desmet, and Gardeniers}}]{SukasLabChip2013}
\bibinfo{author}{\bibfnamefont{S.}~\bibnamefont{Sukas}},
  \bibinfo{author}{\bibfnamefont{R.~M.} \bibnamefont{Tiggelaar}},
  \bibinfo{author}{\bibfnamefont{G.}~\bibnamefont{Desmet}}, \bibnamefont{and}
  \bibinfo{author}{\bibfnamefont{H.~J. G.~E.} \bibnamefont{Gardeniers}},
  \bibinfo{journal}{Lab Chip} \textbf{\bibinfo{volume}{13}},
  \bibinfo{pages}{3061} (\bibinfo{year}{2013}),
  \urlprefix\url{http://dx.doi.org/10.1039/C3LC41311J}.

\bibitem[{\citenamefont{Masoud et~al.}(2013)\citenamefont{Masoud, Khairy, and
  Mousa}}]{MasoudJAC2013}
\bibinfo{author}{\bibfnamefont{E.~M.} \bibnamefont{Masoud}},
  \bibinfo{author}{\bibfnamefont{M.}~\bibnamefont{Khairy}}, \bibnamefont{and}
  \bibinfo{author}{\bibfnamefont{M.}~\bibnamefont{Mousa}}, \bibinfo{journal}{J.
  Alloy. Compd.} \textbf{\bibinfo{volume}{569}}, \bibinfo{pages}{150}
  (\bibinfo{year}{2013}),
  \urlprefix\url{http://dx.doi.org/10.1016/j.jallcom.2013.03.113}.

\bibitem[{\citenamefont{Zardetto et~al.}(2013)\citenamefont{Zardetto, Angelis,
  Vesce, Caratto, Mazzuca, Gasiorowski, Reale, Carlo, and
  Brown}}]{ZardettoNanotech2013}
\bibinfo{author}{\bibfnamefont{V.}~\bibnamefont{Zardetto}},
  \bibinfo{author}{\bibfnamefont{G.~D.} \bibnamefont{Angelis}},
  \bibinfo{author}{\bibfnamefont{L.}~\bibnamefont{Vesce}},
  \bibinfo{author}{\bibfnamefont{V.}~\bibnamefont{Caratto}},
  \bibinfo{author}{\bibfnamefont{C.}~\bibnamefont{Mazzuca}},
  \bibinfo{author}{\bibfnamefont{J.}~\bibnamefont{Gasiorowski}},
  \bibinfo{author}{\bibfnamefont{A.}~\bibnamefont{Reale}},
  \bibinfo{author}{\bibfnamefont{A.~D.} \bibnamefont{Carlo}}, \bibnamefont{and}
  \bibinfo{author}{\bibfnamefont{T.~M.} \bibnamefont{Brown}},
  \bibinfo{journal}{Nanotech.} \textbf{\bibinfo{volume}{24}},
  \bibinfo{pages}{255401} (\bibinfo{year}{2013}),
  \urlprefix\url{http://dx.doi.org/10.1088/0957-4484/24/25/255401}.

\bibitem[{\citenamefont{Zakery and Elliott}(2003)}]{ElliottJNCS2003}
\bibinfo{author}{\bibfnamefont{A.}~\bibnamefont{Zakery}} \bibnamefont{and}
  \bibinfo{author}{\bibfnamefont{S.~R.} \bibnamefont{Elliott}},
  \bibinfo{journal}{J. Non-Cryst. Solids} \textbf{\bibinfo{volume}{330}},
  \bibinfo{pages}{1} (\bibinfo{year}{2003}),
  \urlprefix\url{http://dx.doi.org/10.1016/j.jnoncrysol.2003.08.064}.

\bibitem[{\citenamefont{Vallet-Reg\'{i}
  et~al.}(2003)\citenamefont{Vallet-Reg\'{i}, Ragel, and
  Salinas}}]{RegiEJIC2003}
\bibinfo{author}{\bibfnamefont{M.}~\bibnamefont{Vallet-Reg\'{i}}},
  \bibinfo{author}{\bibfnamefont{C.~V.} \bibnamefont{Ragel}}, \bibnamefont{and}
  \bibinfo{author}{\bibfnamefont{A.~J.} \bibnamefont{Salinas}},
  \bibinfo{journal}{Eur. J. Inorg. Chem.} \textbf{\bibinfo{volume}{2003}},
  \bibinfo{pages}{1029} (\bibinfo{year}{2003}),
  \urlprefix\url{http://dx.doi.org/10.1002/ejic.200390134}.

\bibitem[{\citenamefont{Furtos et~al.}(2013)\citenamefont{Furtos,
  Tomoaia-Cotisel, and Prejmerean}}]{FurtosPST2013}
\bibinfo{author}{\bibfnamefont{G.}~\bibnamefont{Furtos}},
  \bibinfo{author}{\bibfnamefont{M.}~\bibnamefont{Tomoaia-Cotisel}},
  \bibnamefont{and}
  \bibinfo{author}{\bibfnamefont{C.}~\bibnamefont{Prejmerean}},
  \bibinfo{journal}{Particul. Sci. Technol.} \textbf{\bibinfo{volume}{31}},
  \bibinfo{pages}{332} (\bibinfo{year}{2013}),
  \urlprefix\url{http://dx.doi.org/10.1080/02726351.2012.736458}.

\bibitem[{\citenamefont{Gapontsev et~al.}(1982)\citenamefont{Gapontsev,
  Matitsin, Isineev, and Kravchenko}}]{GapontsevOLT1982}
\bibinfo{author}{\bibfnamefont{V.}~\bibnamefont{Gapontsev}},
  \bibinfo{author}{\bibfnamefont{S.}~\bibnamefont{Matitsin}},
  \bibinfo{author}{\bibfnamefont{A.}~\bibnamefont{Isineev}}, \bibnamefont{and}
  \bibinfo{author}{\bibfnamefont{V.}~\bibnamefont{Kravchenko}},
  \bibinfo{journal}{Opt. Laser Technol.} \textbf{\bibinfo{volume}{14}},
  \bibinfo{pages}{184} (\bibinfo{year}{1982}),
  \urlprefix\url{http://dx.doi.org/10.1016/0030-3992(82)90095-0}.

\bibitem[{\citenamefont{Guo et~al.}(2013)\citenamefont{Guo, Zhang, Hu, Chen,
  and Zhang}}]{GuoJLum2013}
\bibinfo{author}{\bibfnamefont{Y.}~\bibnamefont{Guo}},
  \bibinfo{author}{\bibfnamefont{L.}~\bibnamefont{Zhang}},
  \bibinfo{author}{\bibfnamefont{L.}~\bibnamefont{Hu}},
  \bibinfo{author}{\bibfnamefont{N.-K.} \bibnamefont{Chen}}, \bibnamefont{and}
  \bibinfo{author}{\bibfnamefont{J.}~\bibnamefont{Zhang}}, \bibinfo{journal}{J.
  Lum.} \textbf{\bibinfo{volume}{138}}, \bibinfo{pages}{209}
  (\bibinfo{year}{2013}),
  \urlprefix\url{http://dx.doi.org/10.1016/j.jlumin.2013.01.033}.

\bibitem[{\citenamefont{Ghosh et~al.}(2013)\citenamefont{Ghosh, Ghosh, Das,
  Das, and Banerjee}}]{GhoshCPL2013}
\bibinfo{author}{\bibfnamefont{A.}~\bibnamefont{Ghosh}},
  \bibinfo{author}{\bibfnamefont{S.}~\bibnamefont{Ghosh}},
  \bibinfo{author}{\bibfnamefont{S.}~\bibnamefont{Das}},
  \bibinfo{author}{\bibfnamefont{P.~K.} \bibnamefont{Das}}, \bibnamefont{and}
  \bibinfo{author}{\bibfnamefont{R.}~\bibnamefont{Banerjee}},
  \bibinfo{journal}{Chem. Phys. Lett.} \textbf{\bibinfo{volume}{570}},
  \bibinfo{pages}{113} (\bibinfo{year}{2013}),
  \urlprefix\url{http://dx.doi.org/10.1016/j.cplett.2013.03.063}.

\bibitem[{\citenamefont{Carapella et~al.}(2013)\citenamefont{Carapella, Duran,
  Hrdina, Sears, and Tingley}}]{CarapellaJNCS2013}
\bibinfo{author}{\bibfnamefont{A.}~\bibnamefont{Carapella}},
  \bibinfo{author}{\bibfnamefont{C.}~\bibnamefont{Duran}},
  \bibinfo{author}{\bibfnamefont{K.}~\bibnamefont{Hrdina}},
  \bibinfo{author}{\bibfnamefont{D.}~\bibnamefont{Sears}}, \bibnamefont{and}
  \bibinfo{author}{\bibfnamefont{J.}~\bibnamefont{Tingley}},
  \bibinfo{journal}{J. Non-Cryst. Solids} \textbf{\bibinfo{volume}{367}},
  \bibinfo{pages}{37} (\bibinfo{year}{2013}),
  \urlprefix\url{http://dx.doi.org/10.1016/j.jnoncrysol.2013.01.052}.

\bibitem[{\citenamefont{Lee et~al.}(2013)\citenamefont{Lee, Cho, Han, Yoon,
  Lee, Jeong, and Lee}}]{LeeMRB2013}
\bibinfo{author}{\bibfnamefont{S.-M.} \bibnamefont{Lee}},
  \bibinfo{author}{\bibfnamefont{H.-J.} \bibnamefont{Cho}},
  \bibinfo{author}{\bibfnamefont{J.~Y.} \bibnamefont{Han}},
  \bibinfo{author}{\bibfnamefont{H.-J.} \bibnamefont{Yoon}},
  \bibinfo{author}{\bibfnamefont{K.-H.} \bibnamefont{Lee}},
  \bibinfo{author}{\bibfnamefont{D.~H.} \bibnamefont{Jeong}}, \bibnamefont{and}
  \bibinfo{author}{\bibfnamefont{Y.-S.} \bibnamefont{Lee}},
  \bibinfo{journal}{Mater. Res. Bull.} \textbf{\bibinfo{volume}{48}},
  \bibinfo{pages}{1523} (\bibinfo{year}{2013}),
  \urlprefix\url{http://dx.doi.org/10.1016/j.materresbull.2012.12.055}.

\bibitem[{\citenamefont{de~Oliveira et~al.}(2013)\citenamefont{de~Oliveira,
  de~Souza, Dias, de~Carvalho, Mansur, and
  de~Magalh{\~a}es~Pereira}}]{OliveiraBM2013}
\bibinfo{author}{\bibfnamefont{A.~A.~R.} \bibnamefont{de~Oliveira}},
  \bibinfo{author}{\bibfnamefont{D.~A.} \bibnamefont{de~Souza}},
  \bibinfo{author}{\bibfnamefont{L.~L.~S.} \bibnamefont{Dias}},
  \bibinfo{author}{\bibfnamefont{S.~M.} \bibnamefont{de~Carvalho}},
  \bibinfo{author}{\bibfnamefont{H.~S.} \bibnamefont{Mansur}},
  \bibnamefont{and}
  \bibinfo{author}{\bibfnamefont{M.}~\bibnamefont{de~Magalh{\~a}es~Pereira}},
  \bibinfo{journal}{Biomed. Mater.} \textbf{\bibinfo{volume}{8}},
  \bibinfo{pages}{025011} (\bibinfo{year}{2013}),
  \urlprefix\url{http://dx.doi.org/10.1088/1748-6041/8/2/025011}.

\bibitem[{\citenamefont{Weetall and Filbert}(1974)}]{WeetaliME1974}
\bibinfo{author}{\bibfnamefont{H.}~\bibnamefont{Weetall}} \bibnamefont{and}
  \bibinfo{author}{\bibfnamefont{A.}~\bibnamefont{Filbert}},
  \bibinfo{journal}{Methods Enzymol.} \textbf{\bibinfo{volume}{34}},
  \bibinfo{pages}{59} (\bibinfo{year}{1974}),
  \urlprefix\url{http://dx.doi.org/10.1016/S0076-6879(74)34007-4}.

\bibitem[{\citenamefont{Kob and Andersen}(1995)}]{KobAndersenPRE1995}
\bibinfo{author}{\bibfnamefont{W.}~\bibnamefont{Kob}} \bibnamefont{and}
  \bibinfo{author}{\bibfnamefont{H.~C.} \bibnamefont{Andersen}},
  \bibinfo{journal}{Phys. Rev. E} \textbf{\bibinfo{volume}{51}},
  \bibinfo{pages}{4626} (\bibinfo{year}{1995}),
  \urlprefix\url{http://dx.doi.org/10.1103/PhysRevE.51.4626}.

\bibitem[{\citenamefont{Lennard-Jones}(1924)}]{LJProcRSoc1924}
\bibinfo{author}{\bibfnamefont{J.~E.} \bibnamefont{Lennard-Jones}},
  \bibinfo{journal}{Proc. R. Soc. Lond. A} \textbf{\bibinfo{volume}{106}},
  \bibinfo{pages}{463} (\bibinfo{year}{1924}),
  \urlprefix\url{http://dx.doi.org/10.1098/rspa.1924.0081}.

\bibitem[{\citenamefont{Toxvaerd et~al.}(2009)\citenamefont{Toxvaerd, Pedersen,
  Schr{\o}der, and Dyre}}]{UlfJCP2009}
\bibinfo{author}{\bibfnamefont{S.}~\bibnamefont{Toxvaerd}},
  \bibinfo{author}{\bibfnamefont{U.~R.} \bibnamefont{Pedersen}},
  \bibinfo{author}{\bibfnamefont{T.~B.} \bibnamefont{Schr{\o}der}},
  \bibnamefont{and} \bibinfo{author}{\bibfnamefont{J.~C.} \bibnamefont{Dyre}},
  \bibinfo{journal}{J. Chem. Phys.} \textbf{\bibinfo{volume}{130}},
  \bibinfo{pages}{224501} (\bibinfo{year}{2009}),
  \urlprefix\url{http://dx.doi.org/10.1063/1.3144049}.

\bibitem[{\citenamefont{Plimpton}(1995)}]{PlimptonJCompPhys1995}
\bibinfo{author}{\bibfnamefont{S.~J.} \bibnamefont{Plimpton}},
  \bibinfo{journal}{J. Comp. Phys.} \textbf{\bibinfo{volume}{117}},
  \bibinfo{pages}{1} (\bibinfo{year}{1995}),
  \urlprefix\url{http://dx.doi.org/10.1006/jcph.1995.1039}.

\bibitem[{\citenamefont{Swope et~al.}(1982)\citenamefont{Swope, Andersen,
  Berens, and Wilson}}]{SwopeJCP1982}
\bibinfo{author}{\bibfnamefont{W.~C.} \bibnamefont{Swope}},
  \bibinfo{author}{\bibfnamefont{H.~C.} \bibnamefont{Andersen}},
  \bibinfo{author}{\bibfnamefont{P.~H.} \bibnamefont{Berens}},
  \bibnamefont{and} \bibinfo{author}{\bibfnamefont{K.~R.}
  \bibnamefont{Wilson}}, \bibinfo{journal}{J. Chem. Phys.}
  \textbf{\bibinfo{volume}{76}}, \bibinfo{pages}{637} (\bibinfo{year}{1982}),
  \urlprefix\url{http://dx.doi.org/10.1063/1.442716}.

\bibitem[{\citenamefont{Nos\'{e}}(1984)}]{NoseMolPhys1984}
\bibinfo{author}{\bibfnamefont{S.}~\bibnamefont{Nos\'{e}}},
  \bibinfo{journal}{Mol. Phys.} \textbf{\bibinfo{volume}{52}},
  \bibinfo{pages}{255} (\bibinfo{year}{1984}),
  \urlprefix\url{http://dx.doi.org/10.1080/00268978400101201}.

\bibitem[{\citenamefont{Hoover}(1985)}]{HooverPhysRevA1985}
\bibinfo{author}{\bibfnamefont{W.~G.} \bibnamefont{Hoover}},
  \bibinfo{journal}{Phys. Rev. A} \textbf{\bibinfo{volume}{31}},
  \bibinfo{pages}{1695} (\bibinfo{year}{1985}),
  \urlprefix\url{http://dx.doi.org/10.1103/PhysRevA.31.1695}.

\bibitem[{\citenamefont{Morante et~al.}(2006)\citenamefont{Morante, Rossi, and
  Testa}}]{TestaJCP2006}
\bibinfo{author}{\bibfnamefont{S.}~\bibnamefont{Morante}},
  \bibinfo{author}{\bibfnamefont{G.~C.} \bibnamefont{Rossi}}, \bibnamefont{and}
  \bibinfo{author}{\bibfnamefont{M.}~\bibnamefont{Testa}}, \bibinfo{journal}{J.
  Chem. Phys.} \textbf{\bibinfo{volume}{125}}, \bibinfo{pages}{034101}
  (\bibinfo{year}{2006}), \urlprefix\url{http://dx.doi.org/10.1063/1.2214719}.

\bibitem[{\citenamefont{Bitzek et~al.}(2006)\citenamefont{Bitzek, Koskinen,
  G\"{a}hler, Moseler, and Gumbsch}}]{BitzekPRL2006}
\bibinfo{author}{\bibfnamefont{E.}~\bibnamefont{Bitzek}},
  \bibinfo{author}{\bibfnamefont{P.}~\bibnamefont{Koskinen}},
  \bibinfo{author}{\bibfnamefont{F.}~\bibnamefont{G\"{a}hler}},
  \bibinfo{author}{\bibfnamefont{M.}~\bibnamefont{Moseler}}, \bibnamefont{and}
  \bibinfo{author}{\bibfnamefont{P.}~\bibnamefont{Gumbsch}},
  \bibinfo{journal}{Phys. Rev. Lett.} \textbf{\bibinfo{volume}{97}},
  \bibinfo{pages}{170201} (\bibinfo{year}{2006}),
  \urlprefix\url{http://dx.doi.org/10.1103/PhysRevLett.97.170201}.

\bibitem[{\citenamefont{Green}(1954)}]{GreenJCP1954}
\bibinfo{author}{\bibfnamefont{M.~S.} \bibnamefont{Green}},
  \bibinfo{journal}{J. Chem. Phys.} \textbf{\bibinfo{volume}{22}},
  \bibinfo{pages}{398} (\bibinfo{year}{1954}),
  \urlprefix\url{http://dx.doi.org/10.1063/1.1740082}.

\bibitem[{\citenamefont{Kubo}(1957)}]{KuboJPSJ1957}
\bibinfo{author}{\bibfnamefont{R.}~\bibnamefont{Kubo}}, \bibinfo{journal}{J.
  Phys. Soc. Jpn.} \textbf{\bibinfo{volume}{12}}, \bibinfo{pages}{570}
  (\bibinfo{year}{1957}),
  \urlprefix\url{http://dx.doi.org/10.1143/JPSJ.12.570}.

\bibitem[{\citenamefont{Helfand}(1960)}]{HelfandPhysRev1960}
\bibinfo{author}{\bibfnamefont{E.}~\bibnamefont{Helfand}},
  \bibinfo{journal}{Phys. Rev.} \textbf{\bibinfo{volume}{119}},
  \bibinfo{pages}{1} (\bibinfo{year}{1960}),
  \urlprefix\url{http://dx.doi.org/10.1103/PhysRev.119.1}.

\bibitem[{\citenamefont{Viscardy and Gaspard}(2003)}]{GaspardPRE2003}
\bibinfo{author}{\bibfnamefont{S.}~\bibnamefont{Viscardy}} \bibnamefont{and}
  \bibinfo{author}{\bibfnamefont{P.}~\bibnamefont{Gaspard}},
  \bibinfo{journal}{Phys. Rev. E} \textbf{\bibinfo{volume}{68}},
  \bibinfo{pages}{041204} (\bibinfo{year}{2003}),
  \urlprefix\url{http://dx.doi.org/10.1103/PhysRevE.68.041204}.

\bibitem[{\citenamefont{Hummer}(2005)}]{Hummer2005}
\bibinfo{author}{\bibfnamefont{G.}~\bibnamefont{Hummer}}, \bibinfo{journal}{New
  J. Phys.} \textbf{\bibinfo{volume}{7}}, \bibinfo{pages}{34}
  (\bibinfo{year}{2005}),
  \urlprefix\url{http://dx.doi.org/10.1088/1367-2630/7/1/034}.

\bibitem[{\citenamefont{Mittal et~al.}(2006)\citenamefont{Mittal, Errington,
  and Truskett}}]{MittalPRL2006}
\bibinfo{author}{\bibfnamefont{J.}~\bibnamefont{Mittal}},
  \bibinfo{author}{\bibfnamefont{J.~R.} \bibnamefont{Errington}},
  \bibnamefont{and} \bibinfo{author}{\bibfnamefont{T.~M.}
  \bibnamefont{Truskett}}, \bibinfo{journal}{Phys. Rev. Lett.}
  \textbf{\bibinfo{volume}{96}}, \bibinfo{pages}{177804}
  (\bibinfo{year}{2006}),
  \urlprefix\url{http://dx.doi.org/10.1103/PhysRevLett.96.177804}.

\bibitem[{\citenamefont{Sano}(1981)}]{SanoJCP1981}
\bibinfo{author}{\bibfnamefont{H.}~\bibnamefont{Sano}}, \bibinfo{journal}{J.
  Chem. Phys.} \textbf{\bibinfo{volume}{74}}, \bibinfo{pages}{1394}
  (\bibinfo{year}{1981}), \urlprefix\url{http://dx.doi.org/10.1063/1.441203}.

\bibitem[{\citenamefont{Chandrasekhar}(1943)}]{RevModPhys.15.1}
\bibinfo{author}{\bibfnamefont{S.}~\bibnamefont{Chandrasekhar}},
  \bibinfo{journal}{Rev. Mod. Phys.} \textbf{\bibinfo{volume}{15}},
  \bibinfo{pages}{1} (\bibinfo{year}{1943}),
  \urlprefix\url{http://dx.doi.org/10.1103/RevModPhys.15.1}.

\bibitem[{\citenamefont{Lau and Lubensky}(2007)}]{LubenskyPRE2007}
\bibinfo{author}{\bibfnamefont{A.~W.~C.} \bibnamefont{Lau}} \bibnamefont{and}
  \bibinfo{author}{\bibfnamefont{T.~C.} \bibnamefont{Lubensky}},
  \bibinfo{journal}{Phys. Rev. E} \textbf{\bibinfo{volume}{76}},
  \bibinfo{pages}{011123} (\bibinfo{year}{2007}),
  \urlprefix\url{http://dx.doi.org/10.1103/PhysRevE.76.011123}.

\bibitem[{\citenamefont{Perera and Harrowell}(1999)}]{PereraJCP1999}
\bibinfo{author}{\bibfnamefont{D.~N.} \bibnamefont{Perera}} \bibnamefont{and}
  \bibinfo{author}{\bibfnamefont{P.}~\bibnamefont{Harrowell}},
  \bibinfo{journal}{J. Chem. Phys.} \textbf{\bibinfo{volume}{111}},
  \bibinfo{pages}{5441} (\bibinfo{year}{1999}),
  \urlprefix\url{http://dx.doi.org/10.1063/1.479804}.

\bibitem[{\citenamefont{Ashwin and Sastry}(2004)}]{SastryLJ-JPCM2004}
\bibinfo{author}{\bibfnamefont{S.~S.} \bibnamefont{Ashwin}} \bibnamefont{and}
  \bibinfo{author}{\bibfnamefont{S.}~\bibnamefont{Sastry}},
  \bibinfo{journal}{J. Phys.: Condens. Matter} \textbf{\bibinfo{volume}{15}},
  \bibinfo{pages}{S1253} (\bibinfo{year}{2004}),
  \urlprefix\url{http://dx.doi.org/10.1088/0953-8984/15/11/343}.

\bibitem[{\citenamefont{Rault}(1997)}]{RaultVFT1997}
\bibinfo{author}{\bibfnamefont{J.}~\bibnamefont{Rault}}, \bibinfo{journal}{J.
  Non-Cryst. Solids} \textbf{\bibinfo{volume}{271}}, \bibinfo{pages}{177}
  (\bibinfo{year}{1997}),
  \urlprefix\url{http://dx.doi.org/10.1016/S0022-3093(00)00099-5}.

\bibitem[{\citenamefont{Spohr}(1997)}]{SpohrJCP1997}
\bibinfo{author}{\bibfnamefont{E.}~\bibnamefont{Spohr}}, \bibinfo{journal}{J.
  Chem. Phys.} \textbf{\bibinfo{volume}{106}}, \bibinfo{pages}{388}
  (\bibinfo{year}{1997}), \urlprefix\url{http://dx.doi.org/10.1063/1.473202}.

\bibitem[{\citenamefont{Abraham}(1978)}]{AbrahamJCP1978}
\bibinfo{author}{\bibfnamefont{F.~F.} \bibnamefont{Abraham}},
  \bibinfo{journal}{J. Chem. Phys.} \textbf{\bibinfo{volume}{68}},
  \bibinfo{pages}{3713} (\bibinfo{year}{1978}),
  \urlprefix\url{http://dx.doi.org/10.1063/1.436229}.

\bibitem[{\citenamefont{Schofield and
  Henderson}(1982)}]{SchofieldProcRSocLondA1982}
\bibinfo{author}{\bibfnamefont{P.}~\bibnamefont{Schofield}} \bibnamefont{and}
  \bibinfo{author}{\bibfnamefont{J.~R.} \bibnamefont{Henderson}},
  \bibinfo{journal}{Proc. R. Soc. Lond. A} \textbf{\bibinfo{volume}{379}},
  \bibinfo{pages}{231} (\bibinfo{year}{1982}),
  \urlprefix\url{http://dx.doi.org/10.1098/rspa.1982.0015}.

\bibitem[{\citenamefont{Baus and Lovett}(1990)}]{BausPRL1990}
\bibinfo{author}{\bibfnamefont{M.}~\bibnamefont{Baus}} \bibnamefont{and}
  \bibinfo{author}{\bibfnamefont{R.}~\bibnamefont{Lovett}},
  \bibinfo{journal}{Phys. Rev. Lett.} \textbf{\bibinfo{volume}{65}},
  \bibinfo{pages}{1781} (\bibinfo{year}{1990}),
  \urlprefix\url{http://dx.doi.org/10.1103/PhysRevLett.65.1781}.

\bibitem[{\citenamefont{Pehlke and Tersoff}(1991)}]{TersoffPRL1991}
\bibinfo{author}{\bibfnamefont{E.}~\bibnamefont{Pehlke}} \bibnamefont{and}
  \bibinfo{author}{\bibfnamefont{J.}~\bibnamefont{Tersoff}},
  \bibinfo{journal}{Phys. Rev. Lett.} \textbf{\bibinfo{volume}{67}},
  \bibinfo{pages}{465} (\bibinfo{year}{1991}),
  \urlprefix\url{http://dx.doi.org/10.1103/PhysRevLett.67.465}.

\bibitem[{\citenamefont{Blokhuis and Bedeaux}(1992)}]{BlokhuisJCP1992}
\bibinfo{author}{\bibfnamefont{E.~M.} \bibnamefont{Blokhuis}} \bibnamefont{and}
  \bibinfo{author}{\bibfnamefont{D.}~\bibnamefont{Bedeaux}},
  \bibinfo{journal}{J. Chem. Phys.} \textbf{\bibinfo{volume}{97}},
  \bibinfo{pages}{3576} (\bibinfo{year}{1992}),
  \urlprefix\url{http://dx.doi.org/10.1063/1.462992}.

\end{thebibliography}

\end{document}